# Quantitative characterization of the spin orbit torque using harmonic Hall voltage measurements

Masamitsu Hayashi[1] and Junyeon Kim[1]

[1]*National Institute for Materials Science, Tsukuba 305-0047, Japan*

Michihiko Yamanouchi[2,3] and Hideo Ohno[2,3,4]

[2] *Center for Spintronics Integrated Systems, Tohoku University, Sendai 980-8577, Japan*

[3]*Research Institute of Electrical Communication, Tohoku University, Sendai 980-8577, Japan*

[4]*WPI Advanced Institute for Materials Research, Tohoku University, Sendai 980-8577, Japan*

**Solid understanding of current induced torques is key to the development of current and voltage controlled magnetization dynamics in ultrathin magnetic heterostructures. To evaluate the size and direction of such torques, or effective fields, a number of methods have been employed. Here we examine the adiabatic (low frequency) harmonic Hall voltage measurement that has been used to study the effective field. We derive an analytical formula for the harmonic Hall voltages to evaluate the effective field for both out of plane and in-plane magnetized systems. The formula agrees with numerical calculations based on a macrospin model. Two different in-plane magnetized films, Pt|CoFeB|MgO and CuIr|CoFeB|MgO are studied using the formula developed. The effective field obtained for the latter system shows relatively good agreement with that estimated using a spin torque switching phase diagram measurements reported previously. Our results illustrate the versatile applicability of harmonic Hall voltage measurement for studying current induced torques in magnetic heterostructures.**

*Email: hayashi.masamitsu@nims.go.jp

**I. Introduction**

Application of current to systems with large spin orbit coupling in bulk or at interfaces may result in spin current generation via the spin Hall effect[1, 2] and/or current induced spin polarization (the Rashba-Edelstein effect).[3, 4] The generated spin current can act on nearby magnetic moments via spin transfer torque[5, 6] or exchange coupling[7, 8]. These effects are referred to as the "spin orbit torques",[8-14] which is to be distinguished from conventional spin transfer torque since the spin orbit coupling plays a critical role in generating the spin current.

Spin orbit torques are attracting great interest as they can lead to magnetization switching in geometries which were not possible with conventional spin transfer torque,[15, 16] and unprecedented fast domain wall motion[17, 18]. Solid understanding of how these torques arise is thus essential for developing devices utilizing spin orbit effects in ultrathin magnetic heterostructures.

Recently, it has been reported that an adiabatic (low frequency) harmonic Hall voltage measurement scheme, originally developed by Pi *et al.*,[19] can be used to evaluate the "effective magnetic field"[19-21] that generates the torque acting on the magnetic moments.[22-24] This technique has been used to evaluate the size and direction of the effective field in magnetic heterostructures. Using such technique, we have previously shown that the effective field shows a strong dependence on the layer thickness of Ta and CoFeB layers in Ta|CoFeB|MgO heterostructures.[22] The difference in the sign of the spin Hall angle between Ta and Pt has been probed and reported recently.[24] It has also been shown that there is a strong angular dependence (the angle between the magnetization and the current flow direction) of the effective field in Pt|Co|AlOx.[23] Non-local effects, i.e. spin current generated in a Pt layer can propagate through a Cu spacer and exert torques on the magnetic layer, have been probed using a similar technique.[25]

These results show that the adiabatic harmonic Hall voltage measurement is a useful tool to study spin orbit torques in ultrathin magnetic heterostructures.

However, the effective fields evaluated using different techniques vary, in some cases, by order of magnitude.[16, 19-21] It is thus important to examine the accuracy of the characterization method used. With regard to the harmonic Hall measurements, its application to in-plane magnetized systems have been limited, primarily due to the difficulty in obtaining one of the two components (i.e. the spin transfer term) of the effective field. To address these issues, here we derive an analytical formula that describes the harmonic Hall voltages and compare it to numerical calculations based on a macrospin model to first test its validity. We extend our approach, which was previously limited to evaluate out of plane magnetized samples, to characterize in-plane magnetized systems. The formula developed is applied to two different in-plane magnetized systems, Pt|CoFeB|MgO and CuIr|CoFeB|MgO. The effective field obtained for the latter system agrees with that estimated using spin torque switching phase diagrams.

## II. Analytical solutions

### A. Modulation amplitude of the magnetization angle

The magnetic energy of the system can be expressed as

$$E = -K_{EFF} \cos^2 \theta - K_I \sin^2 \varphi \sin^2 \theta - \vec{M} \cdot \vec{H}. \qquad (1)$$

where $K_{EFF}$ is the effective out of plane anisotropy energy and $K_I$ is the in-plane easy axis anisotropy energy densities. $\theta$ and $\varphi$ are the polar and azimuthal angles, respectively, of the magnetization (see Fig. 1 for the definition). $K_{EFF}$ and $K_I$ can be expressed as the following using the demagnetization coefficients $N_i$ ( $\sum_{i=X,Y,Z} N_i = 4\pi$ ) and the uniaxial magnetic anisotropy energy $K_U$:

$$K_{EFF} = K_U - \frac{1}{2}(N_Z - N_X)M_S^2$$
$$K_I = \frac{1}{2}(N_X - N_Y)M_S^2 \qquad (2)$$

$K_U$ is defined positive for out of plane magnetic easy axis. The direction of the magnetization $\vec{M}$ and the external magnetic field $\vec{H}$ are expressed using $\theta$ and $\varphi$ as:

$$\vec{M} = M_S \hat{m}, \quad \hat{m} = (\sin\theta\cos\varphi, \sin\theta\sin\varphi, \cos\theta), \qquad (3)$$

$$\vec{H} = H(\sin\theta_H \cos\varphi_H, \sin\theta_H \sin\varphi_H, \cos\theta_H). \qquad (4)$$

$M_s$ is the saturation magnetization, $\hat{m}$ is a unit vector representing the magnetization direction and $H$ represents the magnitude of the external magnetic field.

The equilibrium magnetization direction ($\theta_0$, $\varphi_0$) is calculated using the following two equations:

$$\frac{\partial E}{\partial \theta} = (K_{EFF} - K_I \sin^2\varphi_0)\sin 2\theta_0 - M_S H(\cos\theta_0 \sin\theta_H (\cos\varphi_0 \cos\varphi_H + \sin\varphi_0 \sin\varphi_H) - \sin\theta_0 \cos\theta_H) = 0, \quad (5a)$$

$$\frac{\partial E}{\partial \varphi} = -K_I \sin^2\theta_0 \sin 2\varphi_0 - M_S H \sin\theta_0 \sin\theta_H \sin(\varphi_H - \varphi_0) = 0. \qquad (5b)$$

Equations (5a) and (5b) can be solved to obtain ($\theta_0$, $\varphi_0$), which will be discussed later. To simplify notations, we define $H_K \equiv \dfrac{2K_{EFF}}{M_S}$ and $H_A \equiv \dfrac{2K_I}{M_S}$.

When current is passed to the device under test, current induced effective field $\Delta H_{X,Y,Z}$, including the Oersted field, can modify the magnetization angle from its equilibrium value ($\theta_0$, $\varphi_0$). The change in the angle, termed the modulation amplitudes $(\Delta\theta, \Delta\varphi)$ hereafter, can be calculated using the following equations:[23, 26]

$$\Delta\theta = \frac{\delta\theta}{\delta H_X}\Delta H_X + \frac{\delta\theta}{\delta H_Y}\Delta H_Y + \frac{\delta\theta}{\delta H_Z}\Delta H_Z, \qquad (8a)$$

$$\Delta\varphi = \frac{\delta\varphi}{\delta H_X}\Delta H_X + \frac{\delta\varphi}{\delta H_Y}\Delta H_Y + \frac{\delta\varphi}{\delta H_Z}\Delta H_Z.  \tag{8b}$$

Here $\frac{\delta\theta}{\delta H_i}$ and $\frac{\delta\varphi}{\delta H_i}$ ($i$=X, Y, Z) represent the degree of change in the angles when a field is applied along one of the (Cartesian) axes. To calculate $\frac{\delta\theta}{\delta H_i}$ and $\frac{\delta\varphi}{\delta H_i}$, we use the following relations derived from Eqs. (5a) and (5b):

$$\frac{\delta}{\delta H_i}\left(\frac{\partial E}{\partial \theta}\right) = 0 = \left[\left(K_{EFF} - K_I \sin^2\varphi_0\right)2\cos 2\theta_0 - M_S\left(-H_X \sin\theta_0 \cos\varphi_0 - H_Y \sin\theta_0 \sin\varphi_0 - H_Z \cos\theta_0\right)\right]\frac{\delta\theta}{\delta H_i}$$
$$+ \left[-K_I \sin 2\theta_0 \sin 2\varphi_0 - M_S \cos\theta_0\left(-H_X \sin\varphi_0 + H_Y \cos\varphi_0\right)\right]\frac{\delta\varphi}{\delta H_i} - M_S f_i \tag{9a}$$

$$\frac{\delta}{\delta H_i}\left(\frac{\partial E}{\partial \varphi}\right) = 0 = \left[-K_I \cos\theta_0 \sin 2\varphi_0 + M_S \cos\theta_0\left(H_X \sin\varphi_0 - H_Y \cos\varphi_0\right)\right]\frac{\delta\theta}{\delta H_i}$$
$$+ \left[-2K_I \sin\theta_0 \cos 2\varphi_0 + M_S \sin\theta_0\left(H_X \cos\varphi_0 + H_Y \sin\varphi_0\right)\right]\frac{\delta\varphi}{\delta H_i} + M_S g_i \tag{9b}$$

where

$$f_i = \begin{bmatrix} \cos\theta_0 \cos\varphi_0 \\ \cos\theta_0 \sin\varphi_0 \\ -\sin\theta_0 \end{bmatrix}, \quad g_i = \begin{bmatrix} \sin\theta_0 \sin\varphi_0 \\ -\sin\theta_0 \cos\varphi_0 \\ 0 \end{bmatrix}.$$

The coupled equations (9a) and (9b) can be solved for $\frac{\delta\theta}{\delta H_i}$ and $\frac{\delta\varphi}{\delta H_i}$, which reads:

$$\frac{\delta\theta}{\delta H_i} = \frac{1}{F_1}\left[f_i - C g_i\right] \tag{10a}$$

$$\frac{\delta\varphi}{\delta H_i} = \frac{1}{F_1 F_2}\left[f_i\left[\frac{1}{2}H_A \sin 2\theta_0 \sin 2\varphi_0 - \cos\theta_0\left(H_X \sin\varphi_0 - H_Y \cos\varphi_0\right)\right] - g_i\left[\left(H_K - H_A \sin^2\varphi_0\right)\cos 2\theta_0 + \vec{H}\cdot\hat{m}\right]\right]$$

(10b)

$$F_1 \equiv \left(H_K - H_A \sin^2\varphi_0\right)\cos 2\theta_0 + \vec{H}\cdot\hat{m} - C\left[\frac{1}{2}H_A \sin 2\theta_0 \sin 2\varphi_0 - \cos\theta_0\left(H_X \sin\varphi_0 - H_Y \cos\varphi_0\right)\right]$$

$$F_2 \equiv -H_A \sin^2\theta_0 \cos 2\varphi_0 + \left(H_X \cos\varphi_0 + H_Y \sin\varphi_0\right)\sin\theta_0$$

$$C \equiv \frac{1}{F_2}\left[\frac{1}{2}H_A \sin 2\theta_0 \sin 2\varphi_0 - \cos\theta_0\left(H_X \sin\varphi_0 - H_Y \cos\varphi_0\right)\right]$$

Substituting Eqs. (10a) and (10b) into Eqs. (8a) and (8b) gives the following expressions for the modulation amplitudes of the magnetization angle:

$$\Delta\theta = \frac{1}{F_1}\left[\left(\cos\theta_0 \cos\varphi_0 - C\sin\theta_0 \sin\varphi_0\right)\Delta H_X + \left(\cos\theta_0 \sin\varphi_0 + C\sin\theta_0 \cos\varphi_0\right)\Delta H_Y - \left(\sin\theta_0\right)\Delta H_Z\right] \tag{11a}$$

$$\Delta\varphi = \frac{1}{F_1 F_2}\left[H_A\left(2\cos^2\theta_0 \cos^2\varphi_0 + \cos 2\theta_0 \sin^2\varphi_0\right) - H_K \cos 2\theta_0 - \vec{H}\cdot\hat{m} - \left(H_X \sin\varphi_0 - H_Y \cos\varphi_0\right)\frac{\cos^2\theta_0 \cos\varphi_0}{\sin\theta_0 \sin\varphi_0}\right]\Delta H_X \sin\theta_0 \sin\varphi_0$$

$$+ \frac{1}{F_1 F_2}\left[H_A\left(2\cos^2\theta_0 \sin^2\varphi_0 - \cos 2\theta_0 \sin^2\varphi_0\right) + H_K \cos 2\theta_0 + \vec{H}\cdot\hat{m} - \left(H_X \sin\varphi_0 - H_Y \cos\varphi_0\right)\frac{\cos^2\theta_0 \sin\varphi_0}{\sin\theta_0 \cos\varphi_0}\right]\Delta H_Y \sin\theta_0 \cos\varphi_0$$

$$+ \frac{1}{F_1 F_2}\left[-\frac{1}{2}H_A \sin 2\theta_0 \sin 2\varphi_0 + \cos\theta_0\left(H_X \sin\varphi_0 - H_Y \cos\varphi_0\right)\right]\Delta H_Z \sin\theta_0$$

(11b)

Equations (11a) and (11b) are valid for any equilibrium magnezation direction and are general for arbitrary values of each parameter (no approximation made). If we assume $|H_A| \ll |H|$, as we do throughout this paper, Eq. (5b) gives $\varphi_0 = \varphi_H$ and thus will simplify Eqs. (11a) and (11b) as the following:

$$\Delta\theta = \frac{\cos\theta_0 \Delta H_{IN} + \sin\theta_0\left(C\Delta\tilde{H}_{IN} - \Delta H_Z\right)}{\left(H_K - H_A \sin^2\varphi_H\right)\cos 2\theta_0 + H\cos(\theta_H - \theta_0) - \frac{1}{2}CH_A \sin 2\theta_0 \sin 2\varphi_H} \tag{11a'}$$

$$\Delta\varphi = \frac{\left[\left(H_K - H_A \sin^2\varphi_H\right)\cos 2\theta_0 + H\cos(\theta_H - \theta_0)\right]\Delta\tilde{H}_{IN} + H_A \sin 2\varphi_H\left[\cos^2\theta_0 \Delta H_{IN} - \frac{1}{2}\sin 2\theta_0 \Delta H_Z\right]}{\left[\left(H_K - H_A \sin^2\varphi_H\right)\cos 2\theta_0 + H\cos(\theta_H - \theta_0) - \frac{1}{2}CH_A \sin 2\theta_0 \sin 2\varphi_H\right]\left[-H_A \sin\theta_0 \cos 2\varphi_H + H\sin\theta_H\right]}$$

(11b')

where $C = \dfrac{H_A \cos\theta_0 \sin 2\varphi_H}{-H_A \sin\theta_0 \cos 2\varphi_H + H\sin\theta_H}$ and we have introduced the following in-plane oriented effective fields

$$\Delta H_{IN} \equiv \Delta H_X \cos\varphi_H + \Delta H_Y \sin\varphi_H$$
$$\Delta\tilde{H}_{IN} \equiv -\Delta H_X \sin\varphi_H + \Delta H_Y \cos\varphi_H \tag{12}$$

## B. Expression for the Hall voltage

The Hall voltage typically contains contributions from the anomalous Hall effect (AHE) and the planar Hall effect (PHE). We define $\Delta R_A$ and $\Delta R_P$ as the change in the Hall resistance due to the AHE and PHE, respectively, when the magnetization direction reverses. Assuming a current flow along the x-axis, the Hall resistance $R_{XY}$ is expressed as:

$$R_{XY} = \frac{1}{2}\Delta R_A \cos\theta + \frac{1}{2}\Delta R_P \sin^2\theta \sin 2\varphi \tag{13}$$

If we substitute $\theta=\theta_0+\Delta\theta$, $\varphi=\varphi_0+\Delta\varphi$ and assume $\Delta\theta\ll1$ and $\Delta\varphi\ll1$, Eq. (13) can be expanded to read

$$R_{XY} \approx \frac{1}{2}\Delta R_A \left(\cos\theta_0 - \Delta\theta\sin\theta_0\right) + \frac{1}{2}\Delta R_P \left(\sin^2\theta_0 + \Delta\theta\sin 2\theta_0\right)\left(\sin 2\varphi_0 + 2\Delta\varphi\cos 2\varphi_0\right) \tag{14}$$

The Hall voltage $V_{XY}$ is a product of the Hall resistance $R_{XY}$ and the current $I$ passed along the device, i.e.

$$V_{XY} = R_{XY} I \tag{15}$$

When a sinusoidal current (excitation amplitude $\Delta I$, frequency $\omega$) is applied, the current induced effective field oscillates in sync with the current. Thus $\Delta H_{X,Y,Z}$, $\Delta H_{IN}$ and $\Delta \tilde{H}_{IN}$ in Eqs. (11a,b) and (11a',b') need to be replaced with $\Delta H_{X,Y,Z}\sin\omega t$, $\Delta H_{IN}\sin\omega t$, and $\Delta \tilde{H}_{IN}\sin\omega t$, respectively. Substituting Eq. (14) into Eq. (15) gives:

$$V_{XY} = V_0 + V_\omega \sin \omega t + V_{2\omega} \cos 2\omega t$$

$$V_0 = \frac{1}{2}(B_\theta + B_\varphi)\Delta I,$$

$$V_\omega = A\Delta I,$$

$$V_{2\omega} = -\frac{1}{2}(B_\theta + B_\varphi)\Delta I \tag{16}$$

$$A = \frac{1}{2}\Delta R_A \cos\theta_0 + \frac{1}{2}\Delta R_P \sin^2\theta_0 \sin 2\varphi_0$$

$$B_\theta = \frac{1}{2}(-\Delta R_A \sin\theta_0 + \Delta R_P \sin 2\theta_0 \sin 2\varphi_0)\Delta\theta$$

$$B_\varphi = \Delta R_P \sin^2\theta_0 \cos 2\varphi_0 \Delta\varphi$$

As evident in Eq. (16), the second harmonic Hall voltage $V_{2\omega}$ contains information of $\Delta H_{X,Y,Z}$ through $\Delta\theta$ and $\Delta\varphi$. Note that Eq. (16) describes the harmonic Hall voltage in the limit of small $\Delta\theta$ and $\Delta\varphi$.

**C. Relation between the current induced effective field and spin torque**

To illustrate the relationship between the current induced effective field $\Delta H_{X,Y,Z}$ and the conventional spin torque terms, $\Delta H_{X,Y,Z}$ can be added, as a vector $\Delta \vec{H}$, in the Landau-Lifshitz-Gilbert (LLG) equation:

$$\frac{\partial \hat{m}}{\partial t} = -\gamma \hat{m} \times \left(-\frac{\partial E}{\partial \vec{M}} + \Delta \vec{H}\right) + \alpha \hat{m} \times \frac{\partial \hat{m}}{\partial t}$$

(17a)

Here $\alpha$ is the Gilbert damping constant, $\gamma$ is the gyromagnetic ratio, $-\frac{\partial E}{\partial \vec{M}}$ is the effective magnetic field that includes external, exchange, anisotropy and demagnetization fields.

Equation (16a) can be compared to the LLG equation with the two spin torque terms.

$$\frac{\partial \hat{m}}{\partial t} = -\gamma \hat{m} \times \left( -\frac{\partial E}{\partial \vec{M}} + a_J (\hat{m} \times \hat{p}) + b_J \hat{p} \right) + \alpha \hat{m} \times \frac{\partial \hat{m}}{\partial t}$$

(17b)

Here $\hat{p}$ represents the magnetization direction of the "reference layer" in spin valve nanopillars/magnetic tunnel junctions, $a_J$ and $b_J$ correspond to the Slonczweski-Berger spin transfer (ST) term[5, 6] and the field-like (FL) term[27], respectively. Comparing Eqs. (17a) and (17b), we can decompose the current induced effective field $\Delta \vec{H}$ into two terms, $\Delta \vec{H} \equiv \Delta \vec{H}_{ST} + \Delta \vec{H}_{FL}$, where $\Delta \vec{H}_{ST} \equiv a_J \hat{m} \times \hat{p}$ and $\Delta \vec{H}_{FL} \equiv b_J \hat{p}$. The spin transfer term $\Delta \vec{H}_{ST}$ depends on the magnetization direction whereas the field-like term $\Delta \vec{H}_{FL}$ is independent of $\hat{m}$. Note that the harmonic Hall voltage measures $\Delta \vec{H}$ and not the torque ($-\gamma (\hat{m} \times \Delta \vec{H})$); thus one can identify whether the effective field is spin transfer-like or field-like by measuring its dependence on the magnetization direction.

For the numerical calculations, we use $\hat{p} = (0,1,0)$ as this represents the spin direction of the electrons entering the CoFeB layer via the spin Hall effect in Ta when current is passed along the +x axis for Ta|CoFeB|MgO heterostructures. Note that the directions of the x- and y- axes are defined differently from that in our previous report.[22] In the following, we consider two representative cases, systems with out of plane and in-plane magnetization.

### III. Approximate expressions for the harmonic Hall voltages
#### A. Out of plane magnetization systems

We first consider a system where the magnetization points along the film normal owing to its perpendicular magnetic anisotropy. To obtain analytical solutions for the harmonic Hall voltages, we make several approximations. First, to solve Eqs. (5a) and (5b) analytically, we assume that the in-plane uniaxial anisotropy is small compared to the external field, i.e. $|H_A| \ll |H|$. Equation (5b) then gives $\varphi_0 = \varphi_H$. Next we assume that the equilibrium magnetization direction does not deviate much from the z-axis, i.e. $\theta_0 = \theta_0'$ ($\theta_0' \ll 1$) for $\vec{M}$ along $+\hat{z}$ and $\theta_0 = \pi - \theta_0'$ ($\theta_0' \ll 1$) for $\vec{M}$ along $-\hat{z}$. Keeping terms that are linear with $\theta_0'$, Eqs. (5a) and (5b) give:

$$\theta_0' = \frac{H \sin \theta_H}{(H_K - H_A \sin^2 \varphi_H) \pm H \cos \theta_H}, \quad \varphi_0 = \varphi_H \tag{18}$$

The $\pm$ sign corresponds to the case for $\vec{M}$ pointing along $\pm\hat{z}$. Assuming $\theta_0' \ll 1$ (keep terms that are linear with $\theta_0'$) expressions (11a') and (11b') can be simplified to read:

$$\Delta\theta \approx \frac{\pm\Delta H_{IN} + \theta_0' \left( C\Delta\tilde{H}_{IN} - \Delta H_Z \right)}{H_K - H_A \sin^2 \varphi_H \pm H(\cos\theta_H + \theta_0' \sin\theta_H) \mp C\theta_0' H_A \sin 2\varphi_H} \tag{19a}$$

$$\Delta\varphi \approx \frac{\left(H_K - H_A \sin^2\varphi_H \pm H(\cos\theta_H + \theta_0'\sin\theta_H)\right)\Delta\tilde{H}_{IN} + H_A \sin 2\varphi_H \left(\Delta H_{IN} \mp \theta_0' \Delta H_Z\right)}{\left[H_K - H_A \sin^2\varphi_H \pm H(\cos\theta_H + \theta_0'\sin\theta_H) \mp CH_A \sin 2\varphi_H\right]\left[-H_A \theta_0' \cos 2\varphi_H + H \sin\theta_H\right]}$$
(19b)

Note that $C \approx \dfrac{\pm H_A \sin 2\varphi_H}{-H_A \theta_0' \cos 2\varphi_H + H \sin\theta_H}$ and the $\pm$ sign stands for $\vec{M}$ along $\pm\hat{z}$ axis. If we consider cases where the external field is directed along one of the Cartesian coordinate axes (along x- or y-axis), then terms with $\sin(2\varphi_H)$ will vanish ($C$ becomes zero). Equations (19a) and (19b) are simplified as:

$$\Delta\theta \approx \frac{\pm\Delta H_{IN} - \theta_0' \Delta H_Z}{H_K - H_A \sin^2\varphi_H \pm H(\cos\theta_H + \theta_0'\sin\theta_H)} \tag{19a'}$$

$$\Delta\varphi \approx \frac{\Delta\tilde{H}_{IN}}{-H_A\theta'_0\cos 2\varphi_H + H\sin\theta_H} \qquad (19b')$$

Substituting Eqs. (19a') and (19b') into Eq. (16) gives:

$$V_\omega \approx \pm\frac{1}{2}\Delta R_A\left[1-\frac{1}{2}\theta'^2_0\right]\Delta I \qquad (20a)$$

$$V_{2\omega} \approx -\frac{1}{2}\left[-\frac{1}{2}\Delta R_A\frac{\pm\Delta H_{IN}}{H_K - H_A\sin^2\varphi_H \pm H\cos\theta_H} + \Delta R_P\frac{\Delta\tilde{H}_{IN}\cos 2\varphi_H}{H_K - H_A\sin^2\varphi_H \pm H\cos\theta_H - H_A\cos 2\varphi_H}\right]\theta'_0\Delta I$$
(20b)

where we have kept the second order (quadratic) term with $\theta'_0$ in Eq. (20a) since the first order (linear) term vanishes. Substituting Eq. (17) into Eqs. (20a) and (20b) and assuming $|H_A|<<|H_K|$, we obtain:

$$V_\omega \approx \pm\frac{1}{2}\Delta R_A\left[1-\frac{1}{2}\left(\frac{H\sin\theta_H}{H_K \pm H\cos\theta_H}\right)^2\right]\Delta I \qquad (21a)$$

$$V_{2\omega} \approx -\frac{1}{4}\left[\mp\Delta R_A\Delta H_{IN} + 2\Delta R_P\Delta\tilde{H}_{IN}\cos 2\varphi_H\right]\frac{H\sin\theta_H}{(H_K \pm H\cos\theta_H)^2}\Delta I \qquad (21b)$$

Equation (21b) shows that the planar Hall effect mixes the signal from different components of the current induced effective field.[23] For systems with negligible PHE, $\Delta H_X$ ($\Delta H_Y$) can be determined by measuring $V_{2\omega}$ as a function of the external in-plane field directed along the x- (y-) axis. However, if the PHE becomes comparable to the size of AHE, contribution from the orthogonal component appears in $V_{2\omega}$ via the PHE. When $\Delta R_P$ is larger than half of $\Delta R_A$, $\Delta H_Y$ ($\Delta H_X$) becomes the dominant term in $V_{2\omega}$ for field sweep along the x- (y-) axis. Thus to estimate the effective field components accurately in systems with non-negligble PHE, one needs to measure $V_{2\omega}$ in two orthogonal directions and analytically calculate each component, as described below.

We follow the procedure used previously[22] to eliminate the prefactors that are functions of $\Delta I$ and $H_K$ in Eqs. (21a) and (21b). $\theta_H = \pi/2$ is substituted in Eqs. (21a) and (21b) since the external field is swept along the film plane. The respective curvature and slope of $V_\omega$ and $V_{2\omega}$ versus the external field are calculated to obtain the ratio $B$:

$$B \equiv \left( \frac{\partial V_{2\omega}}{\partial H} \bigg/ \frac{\partial^2 V_\omega}{\partial H^2} \right) = -\frac{1}{2}\left[ \left( \Delta H_X \mp 2\frac{\Delta R_P}{\Delta R_A}\cos 2\varphi_H \Delta H_Y \right)\cos\varphi_H + \left( \Delta H_Y \pm 2\frac{\Delta R_P}{\Delta R_A}\cos 2\varphi_H \Delta H_X \right)\sin\varphi_H \right] \quad (22)$$

We define $B_X \equiv \left( \frac{\partial V_{2\omega}}{\partial H} \bigg/ \frac{\partial^2 V_\omega}{\partial H^2} \right)\bigg|_{\varphi_H=0}$ and $B_Y \equiv \left( \frac{\partial V_{2\omega}}{\partial H} \bigg/ \frac{\partial^2 V_\omega}{\partial H^2} \right)\bigg|_{\varphi_H=\frac{\pi}{2}}$, which correspond to $B$ measured for $\varphi_H=0$ and $\varphi_H=\pi/2$, respectively, and $\xi \equiv \frac{\Delta R_P}{\Delta R_A}$, which is the ratio of the PHE and AHE resistances. Finally, we obtain:

$$\begin{aligned} \Delta H_X &= -2\frac{(B_X \pm 2\xi B_Y)}{1-4\xi^2} \\ \Delta H_Y &= -2\frac{(B_Y \pm 2\xi B_X)}{1-4\xi^2} \end{aligned} \quad (23)$$

Equation (23) provides a simple method to obtain the effective field under circumstances where both AHE and PHE contribute to the Hall signal. When PHE is negligible, $\xi=0$ and we recover the form derived previously.[22]

### B. In-plane magnetization systems

We next consider systems with in-plane magnetization (easy axis is along the x-axis). We make the same approximation used in the previous section, $|H_A|<<|H|, |H_K|$ (note that $H_K<0$ for in-plane magnetized systems). This may not apply to systems where the shape anisotropy is large, e.g. in high aspect ratio elements such as nanowires. However, lifting this restriction will introduce more complexity in the analysis of the harmonic voltage, in particular, because one cannot make the assumption $\varphi_0 \sim \varphi_H$. Here we limit systems with small $H_A$.

From the discussions on out of plane magnetized systems, the external field must be swept along two directions orthogonal to the magnetization direction in order to obtain each component of the effective field (the spin transfer (ST) and the field-like (FL) terms). It turns out that for the in-plane magnetized systems, the planar Hall contribution becomes larger than that of the anomalous Hall effect, and thus the out of plane field sweep gives information of the FL term whereas the in-plane transverse field sweep (transverse to the magnetization direction) provides the ST term. However, the assumption $|H_A|<<|H|$ used here causes the magnetization to rotate along the transverse field direction as soon as $|H|$ is larger than $|H_A|$, thus hindering evaluation of the ST term. To circumvent this difficulty, we show two alternative approaches. One is to make use of the anisotropic magnetoresistance (AMR) of the magnetic material and measure the longitudinal voltage ($V_{XX}$), which in turn provides information of the ST term. The external field is swept along the film normal but must be kept small to maintain $\theta_0' \ll 1$ when $\theta_0 = \dfrac{\pi}{2} \pm \theta_0'$.

During the field sweep, one can measure $V_{XY}$ and $V_{XX}$ to obtain the FL and ST terms, respectively. The second approach is to sweep the external field ($H$) again along the film normal but in a larger field range, i.e. $|H|$ must be larger than the anisotropy field ($|H_K|$). Here, the second harmonic voltage at low field provides information on the FL term, as before, but the high field value (when $|H|\sim\pm|H_K|$) gives information on the ST term. We start from the small field limit with simultaneous measurements of Hall and longitudinal voltage measurements, followed by the large range field sweep measurements.

### 1. Small field limit

#### (a) Harmonic Hall voltage and the field-like term

We assume that the equilibrium magnetization direction does not deviate much from the film plane, i.e. $\theta_0 = \frac{\pi}{2} + \theta_0'$ ($\theta_0' \ll 1$) for $\vec{M}$ pointing along one direction within the film plane and $\theta_0 = \frac{\pi}{2} - \theta_0'$ ($\theta_0' \ll 1$) for the case where the direction of $\vec{M}$ is reversed from the former. Note that $\varphi_0 = \varphi_H$ for $|H_A| \ll |H|$ (see Eq. (5b)). Keeping terms that are linear with $\theta_0'$, Eqs. (5a) and (5b) give:

$$\theta_0' = \pm \frac{H \cos \theta_H}{\left(H_K - H_A \sin^2 \varphi_H\right) - H \sin \theta_H}, \quad \varphi_0 = \varphi_H. \tag{24}$$

The $\pm$ sign corresponds to the case for $\theta_0 = \theta_0' \pm \frac{\pi}{2}$. Assuming $\theta_0' \ll 1$ and $|H_A| \ll |H|$, $|H_K|$, expressions (11a) and (11b) can be simplified to read:

$$\Delta\theta \approx \frac{\mp \theta_0 \Delta H_{IN} + \left(C\Delta\tilde{H}_{IN} - \Delta H_Z\right)}{-\left(H_K - H_A \sin^2 \varphi_H\right) + H\left(\sin \theta_H \mp \theta_0' \cos \theta_H\right) \pm \theta_0' C H_A \sin 2\varphi_H} \tag{25a}$$

$$\Delta\varphi \approx \frac{\left[-\left(H_K - H_A \sin^2 \varphi_H\right) + H\left(\sin \theta_H \mp \theta_0' \cos \theta_H\right)\right]\Delta\tilde{H}_{IN} + H_A \sin 2\varphi_H \left[\theta_0'^2 \Delta H_{IN} \pm \theta_0' \Delta H_Z\right]}{\left[-\left(H_K - H_A \sin^2 \varphi_H\right) + H\left(\sin \theta_H \mp \theta_0' \cos \theta_H\right) \pm \theta_0' C H_A \sin 2\varphi_H\right]\left[-H_A \cos 2\varphi_H + H \sin \theta_H\right]}$$

(25b)

Again assuming that the external field is directed along one of the Cartesian coordinate axes (along x- or y-axis), terms with $\sin(2\varphi_H)$ can be dropped off. Equations (25a) and (25b) are then simplified as:

$$\Delta\theta \approx \frac{\mp\theta_0'\Delta H_{IN} - \Delta H_Z}{-(H_K - H_A \sin^2\varphi_H) + H(\sin\theta_H \mp \theta_0'\cos\theta_H)} \tag{25a'}$$

$$\Delta\varphi \approx \frac{\Delta\tilde{H}_{IN}}{-H_A \cos 2\varphi_H + H\sin\theta_H} \tag{25b'}$$

Substituting Eqs. (24), (25a') and (25b') into Eq. (16) gives:

$$V_\omega \approx -\frac{1}{2}\Delta R_A \frac{H\cos\theta_H}{(H_K - H_A \sin^2\varphi_H) - H\sin\theta_H} \Delta I \tag{26}$$

$$V_{2\omega} \approx -\frac{1}{2}\left[\frac{1}{2}\Delta R_A \frac{\Delta H_Z}{-(H_K - H_A \sin^2\varphi_H) + H\sin\theta_H} + \Delta R_P \frac{\Delta\tilde{H}_{IN}\cos 2\varphi_H}{-H_A \cos 2\varphi_H + H\sin\theta_H}\right]\Delta I$$

(27)

The second term (the $\Delta R_P$ term) in Eq. (27) dominates the second harmonic voltage for samples with large $|H_K|$. Thus for a typical in-plane magnetized samples, one can ignore the first term in Eq. (27) if the external field is small (i.e. $|H|<<|H_K|$) to obtain:

$$V_{2\omega} \approx -\frac{1}{2}\left[\Delta R_P \frac{\Delta\tilde{H}_{IN}\cos 2\varphi_H}{-H_A \cos 2\varphi_H + H\sin\theta_H}\right]\Delta I \tag{28}$$

It should be noted that the sign of $V_{2\omega}$ at a given $H$, which determines the sign of the effective field, changes when the field is slightly tilted in one direction or the other (e.g. $\theta_H$~1 or −1 deg). When the field is directed exactly along the film normal ($\theta_H$=0), the field dependence of $V_{2\omega}$ vanishes.

When the magnetization and the current flow directions are along +x and the incoming electrons' spin polarization is set to +y ($\hat{p} = (0,1,0)$), the ST *effective field* is directed along the film normal (not to be confused with the ST *torque* that points along the film plane). That is, the ST term corresponds to $\Delta H_Z$ and $\Delta H_X$ is zero. See the inset of Fig. 3(b) in which we show the three effective field components $\Delta H_{X,Y,Z}$ for this geometry. Thus Eq. (28) only allows evaluation of the FL term, i.e. $\Delta H_Y$.

Using Eqs. (26) and (28), we derive a simple formula, similar to that shown in Eq. (23), to estimate $\Delta H_Y$. If the external field is slightly tilted from the film normal, it is preferable to apply the in-plane field component along the magnetic easy axis ($\varphi_H=0$ deg) to unambiguously set the equilibrium magnetization azimuthal angle $\varphi_0$. Substituting $\varphi_H=0$ in Eqs. (26a) and (28) gives:

$$\frac{\partial V_\omega}{\partial H} \approx -\frac{1}{2}\Delta R_A \Delta I \frac{\cos\theta_H}{H_K} \tag{29a}$$

$$\frac{\partial(1/V_{2\omega})}{\partial H} \approx -\frac{2\sin\theta_H}{\Delta R_P \Delta I \Delta H_Y} \tag{29b}$$

Thus $\Delta H_Y$ (the field-like term) can be obtained by:

$$\Delta H_Y \approx \frac{\sin 2\theta_H}{2\xi H_K}\left[1\bigg/\left(\frac{\partial V_\omega}{\partial H}\right)\left(\frac{\partial(1/V_{2\omega})}{\partial H}\right)\right] \tag{30}$$

Unlike the case derived in the previous section (Eq. (23)) here one needs to substitute $H_K$ and $\theta_H$ to calculate the effective field. This is because the field dependence of the first harmonic voltage is primarily determined by the change in the magnetization direction along the z-axis (relevant anisotropy is $H_K$) whereas that of the second harmonic voltage is dominated by the magnetization angular change within the film plane (relevant anisotropy is $H_A$); therefore, taking the ratio of the two harmonic voltages will not cancel out $H_K$ (and $\theta_H$).

### (b) Harmonic longitudinal voltage and the ST term

As noted above, in order to obtain the ST term, one can make use of the AMR effect, if any, of the magnetic material. The longitudinal resistance $R_{XX}$ is expressed as:

$$R_{XX} = R_0 + \frac{1}{2}\Delta R_{MR} \sin^2\theta \cos^2\varphi \tag{31}$$

where $R_0$ and $\Delta R_{MR}$ are, respectively, the resistance independent of the magnetization direction and the change in the resistance due to the AMR effect. Current is assumed to flow along the x-axis. We substitute $\theta=\theta_0+\Delta\theta$, $\varphi=\varphi_0+\Delta\varphi$ into Eq. (31) and assume $\Delta\theta\ll1$ and $\Delta\varphi\ll1$, which then reads:

$$R_{XX} \approx R_0 + \frac{1}{2}\Delta R_{MR}\left(\sin^2\theta_0\cos^2\varphi_0 + \Delta\theta\sin 2\theta_0\cos^2\varphi_0 - \Delta\varphi\sin^2\theta_0\sin 2\varphi_0 - (\Delta\theta^2+\Delta\varphi^2)\sin^2\theta_0\cos^2\varphi_0\right)$$
(32)

Here, we have kept the 2nd order terms that scales with $\Delta\theta^2$ and $\Delta\varphi^2$ to show that these terms cannot be neglected when the current induced effective field $\Delta H_{X,Y,Z}$ (that determines the magnitude of $\Delta\theta$ and $\Delta\varphi$) becomes larger. This is because, for a typical geometry that would be employed here (current flow and the equilibrium magnetization directions pointing along +x), $\sin(2\theta_0)\sim0$ and $\sin(2\varphi_0)\sim0$, and thus the linear terms (that scales with $\Delta\theta$ and $\Delta\varphi$) can be smaller than the 2nd order terms when $\Delta\theta$ or $\Delta\varphi$ is larger than a critical value. As a consequence, there is a limit in $\Delta H_{X,Y,Z}$ above which we cannot neglect the 2nd order terms and this limit is much smaller than the other geometries described previously. For simplicity, we only consider the small limit of $\Delta H_{X,Y,Z}$ here.

Again, for in-plane magnetized systems, we substitute $\theta_0 = \frac{\pi}{2}+\theta_0'$ ($\theta_0'\ll1$) for $\vec{M}$ pointing along one direction within the film plane and $\theta_0 = \frac{\pi}{2}-\theta_0'$ ($\theta_0'\ll1$) when $\vec{M}$ is reversed. With $|H_A|\ll|H|$, $\varphi\sim\varphi_H$, Eq. (32) reads:

$$R_{XX} \approx R_0 + \frac{1}{2}\Delta R_{MR}\left(\cos^2\theta_0' \cos^2\varphi_H \mp \Delta\theta \sin 2\theta_0' \cos^2\varphi_H - \Delta\varphi \cos^2\theta_0 \sin 2\varphi_H\right) \quad (33)$$

Application of a sinusoidal excitation current $I = \Delta I \sin\omega t$ results in a longitudinal voltage which is expressed as:

$$V_{XX} = R_{XX}I = R_{XX}\Delta I \sin\omega t \quad (34)$$

Assuming that the current induced effective fields are in sync with the excitation current, Eqs. (33) and (34) give:

$$
\begin{aligned}
V_{XX} &= V_0^{XX} + V_\omega^{XX}\sin\omega t + V_{2\omega}^{XX}\cos 2\omega t \\
V_0^{XX} &= \frac{1}{2}\tilde{B}\Delta I \\
V_\omega^{XX} &= \tilde{A}\Delta I \\
V_{2\omega}^{XX} &= -\frac{1}{2}\tilde{B}\Delta I \\
\tilde{A} &= R_0 + \frac{1}{2}\Delta R_{MR}\cos^2\theta_0'\cos^2\varphi_H \\
\tilde{B} &= \frac{1}{2}\Delta R_{MR}\left[\mp\Delta\theta\sin 2\theta_0'\cos^2\varphi_H - \Delta\varphi\cos^2\theta_0\sin 2\varphi_H\right]
\end{aligned}
\quad (35)$$

Keeping terms the lowest order terms with $\theta_0'$, we obtain

$$V_\omega^{XX} = \left[R_0 + \frac{1}{2}\Delta R_{MR}\cos^2\varphi_H\left(1-\theta_0'^2\right)\right]\Delta I \quad (36a)$$

$$V_{2\omega}^{XX} = -\frac{1}{4}\Delta R_{MR}\left[\mp 2\theta_0'\Delta\theta\cos^2\varphi_H - \Delta\varphi\sin 2\varphi_H\right]\Delta I \quad (36b)$$

Substituting Eqs. (25a') and (25b') into Eq. (36b) results in the following expression:

$$V_{2\omega}^{XX} = -\frac{1}{4}\Delta R_{MR}\left[\mp 2\theta_0'\left(\frac{\Delta H_Z}{H_K - H_A\sin^2\varphi_H - H\sin\theta_H}\right)\cos^2\varphi_H - \frac{\Delta\tilde{H}_{IN}}{-H_A\cos 2\varphi_H + H\sin\theta_H}\sin 2\varphi_H\right]\Delta I$$
(36b')

where we ignored higher order terms of $\theta_0'$. If one assumes that the external field is directed along one of the Cartesian coordinate axes (along x- or y-axis), terms with $\sin(2\varphi_H)$ can be dropped off. Substituting Eq. (24) into Eqs. (36a) and (36b') results in:

$$V_\omega^{XX} = \left[ R_0 + \frac{1}{2}\Delta R_{MR} \cos^2 \varphi_H \left( 1 - \left[ \frac{H \cos \theta_H}{H_K - H_A \sin^2 \varphi_H - H \sin \theta_H} \right]^2 \right) \right] \Delta I \quad (37a)$$

$$V_{2\omega}^{XX} = \frac{1}{2}\Delta R_{MR} \cos^2 \varphi_H \left[ \frac{H \cos \theta_H \Delta H_Z}{\left(H_K - H_A \sin^2 \varphi_H - H \sin \theta_H\right)^2} \right] \Delta I \quad (37b)$$

These expressions are similar to those of Eqs. (21a) and (21b). If we consider that the external field is directed along the z-axis, i.e. $\theta_H \ll 1$, then the ratio of the field derivatives of the first and second harmonic signals directly provides the ST term ($\Delta H_Z$):

$$\Delta H_Z = -2 \left( \frac{\partial V_{2\omega}^{XX}}{\partial H} \bigg/ \frac{\partial^2 V_\omega^{XX}}{\partial H^2} \right) \quad (38)$$

Equations (30) and (38) show that combination of the Hall and longitudinal voltage measurements can provide means to evaluate both components of the effective field for in-plane magnetized samples.

### 2. Large field sweep

In material systems with small anisotropic magnetoresistance (AMR), it is difficult to evaluate the ST term for in-plane magnetized materials. To overcome this problem, one can sweep the external field along the film normal and estimate the FL and ST terms from the low and high field regimes, respectively. This is a combination of the methods described in previous sections: the low field regime is discussed in section IIIB.1(a) whereas the high field regime is noted in section IIIA (the large out of plane field forces the moments to point along the film normal and thus the geometry becomes similar to that of the out of plane magnetized system). So far we have assumed that the equilibrium magnetization polar angle $\theta_0$ is close to either along the film plane or normal to it and its change due to the external field application is small. This assumption allowed us to derive approximate expressions for the first and second harmonic voltage signals. If $\theta_0$ varies with the field in a large way, such approximation cannot be made. Thus here we instead use the first harmonic Hall voltage ($V_\omega$) to estimate $\theta_0$ at each applied field and substitute it into the second harmonic Hall voltage ($V_{2\omega}$) expression.

To simplify notations, we assume $\varphi_0 \sim \varphi_H$ and consider cases where $\varphi_H$ points along the x or y axis. Equation (16) then reduces to the following expression:

$$V_{XY} = V_0 + V_\omega \sin \omega t + V_{2\omega} \cos 2\omega t$$

$$V_0 = \frac{1}{2}\left(\bar{B}_\theta + \bar{B}_\varphi\right)\Delta I,$$

$$V_\omega = \bar{A}\Delta I,$$

$$V_{2\omega} = -\frac{1}{2}\left(\bar{B}_\theta + \bar{B}_\varphi\right)\Delta I$$

$$\bar{A} = \frac{1}{2}\Delta R_A \cos\theta_0 \tag{39}$$

$$\bar{B}_\theta = \frac{1}{2}\frac{\Delta R_A \sin\theta_0 \left(\Delta H_Z \sin\theta_0 - \Delta H_{IN}\cos\theta_0\right)}{\left(H_K - H_A \sin^2\varphi_H\right)\cos 2\theta_0 + H\cos(\theta_H - \theta_0)}$$

$$\bar{B}_\varphi = \frac{\Delta R_P \Delta \tilde{H}_{IN} \sin^2\theta_0 \cos 2\varphi_H}{-H_A \sin\theta_0 \cos 2\varphi_H + H\sin\theta_H}$$

It turns out that for $\hat{p} = (0, \pm 1, 0)$, it is sufficient to measure the harmonic signals along the x-direction, i.e. $\varphi_H=0$ or $\pi$. Then $V_\omega$ and $V_{2\omega}$ can be simplified as:

$$V_\omega = \frac{1}{2}\Delta R_A \cos\theta_0 \Delta I \tag{40a}$$

$$V_{2\omega} = -\frac{\Delta R_A}{2}\left(\frac{1}{2}\frac{\sin\theta_0 \cos\varphi_H}{H_K \cos 2\theta_0 + H\cos(\theta_H - \theta_0)}a_J + \xi\frac{\sin^2\theta_0 \cos\varphi_H}{-H_A \sin\theta_0 + H\sin\theta_H}b_J\right)\Delta I \tag{40b}$$

where we have used $a_J$ and $b_J$ from Eq. (16b) and $\xi \equiv \frac{\Delta R_P}{\Delta R_A}$ as before. To obtain Eq. (40b), we first note that:

$$\Delta \vec{H} = (\Delta H_X, \Delta H_Y, \Delta H_Z) = (-a_J \cos\theta_0,\ b_J,\ a_J \sin\theta_0 \cos\varphi_H) \quad \{\text{for } \hat{p} = (0,1,0)\} \tag{41}$$

Using Eq. (41), the expression inside the bracket of the numerator in $\bar{B}_\theta$ (Eq. (39)) can be simplified, that is, $\Delta H_Z \sin\theta_0 - \Delta H_X \cos\varphi_H \cos\theta_0 = a_J \cos\varphi_H$ (we assume $\varphi_H = 0$ or $\pi$ which gives $\Delta H_{IN} = \Delta H_X \cos\varphi_H$). Substituting these expressions into Eq. (39) yields Eq. (40b). Equation (40b) takes a form similar to that of Eq. (27): the first term depends on $a_J$ and $\Delta R_A$ but its contribution is small unless $H \sim |H_K|$ or larger. The second term is proportional to $b_J$ and $\Delta R_P$ and dominates $V_{2\omega}$ at low fields. Thus the low field values of $V_{2\omega}$ provide information of $b_J$, as discussed in section IIIB.1(a) (e.g. Eq. (30)), whereas the higher field values give $a_J$.

To estimate $a_J$ and $b_J$, we use the measured $V_\omega$ and Eq. (40a) to evaluate $\theta_0$ at each $H$. The field dependence of $V_\omega$ also provides value of $\Delta R_A \Delta I$ and information on $H_K$. $H_K$ can be roughly estimated by extrapolating a linear fitting of $V_\omega$ vs. $H$ near zero field to the saturation value of $V_\omega$ at large fields (the intersection field is $\sim H_K$), as commonly done in estimating $H_K$ using magnetization hysteresis loops. At each $H$, one can substitute $\theta_0$ into Eq. (40b) and fit $V_{2\omega}$ vs. $H$ with $a_J$ and $b_J$ as the fitting parameters. It should be noted that when the field direction is reversed in one field sweep, it is not appropriate to substitute a 'negative $H$' in Eq. (40b); rather one needs to change $\theta_0, \theta_H, \varphi_H$ to $\pi-\theta_0, \pi-\theta_H, \varphi_H+\pi$, respectively and keep $H$ positive. In effect, this will result in changing the sign of $V_{2\omega}$ for negative fields.

## IV. Comparison to numerical calculations

The analytical solutions derived above are compared to numerical calculations. We solve Eq. (17b) numerically to obtain the equilibrium magnetization direction and the associated harmonic voltage signals. A macrospin model[28, 29] is used to describe the system. Substituting Eq. (1) into Eq. (17b), the following differential equations are obtained:

$$\frac{1+\alpha^2}{\gamma}\frac{\partial \theta}{\partial t} = \alpha h_\theta + h_\varphi$$

$$\frac{1+\alpha^2}{\gamma}\frac{\partial \varphi}{\partial t}\sin\theta = -h_\theta + \alpha h_\varphi$$

$$h_\theta = a_J \cos\varphi + b_J \cos\theta \sin\varphi + h_X \cos\theta \cos\varphi + h_Y \cos\theta \sin\varphi - h_Z \sin\theta$$

$$h_\varphi = a_J \cos\theta \sin\varphi + b_J \cos\varphi - h_X \sin\varphi + h_Y \cos\varphi \qquad (42)$$

$$h_X = H_X - M_S N_X \sin\theta \cos\varphi$$

$$h_Y = H_Y - M_S N_Y \sin\theta \sin\varphi$$

$$h_Z = H_Z - (M_S N_Z - H_K)\cos\theta$$

The two coupled differential equations are numerically solved to obtain the equilibrium magnetization direction when both the external and the current induced effective fields are turned on. To mimic the experimental setup, a sinusoidal current is passed along the x-axis and the resulting Hall and longitudinal voltages are evaluated. Contributions from the anomalous Hall effect (AHE) and the planar Hall effect (PHE) are considered for the Hall voltage and that from the anisotropic magnetoresistance (AMR) is taken into account for the longitudinal voltage. For a given time during one cycle of the sinusoidal current application, we calculate the equilibrium magnetization direction and the corresponding Hall and longitudinal voltages. One cycle is divided into two hundred time steps to obtain the temporal variation of the Hall and longitudinal voltages. The calculated voltages are fitted with Eq. (16): i.e. $V_{XY} = V_0 + V_\omega \sin\omega t + V_{2\omega}\cos 2\omega t$, to obtain the first and second harmonic signals. We compare the numerical results with the analytical solutions derived in the previous sections.

### A. Out of plane magnetization systems

Figure 2 shows results for the out of plane magnetized samples. The equilibrium magnetization angle ($\theta_0$) with respect to the film normal (a), first (b) and second (d) harmonic Hall voltages are plotted against an in-plane field directed along the current flow direction (i.e. along the x-axis). The transverse field (directed along the y-axis) dependence of the second harmonic Hall voltage is shown in Fig. 2(f); the corresponding magnetization angle ($\theta_0$) and the first harmonic Hall voltage are the same with those shown in (a) and (b), respectively. The parameters used here mimic the system of Ta|CoFeB|MgO heterostructures[22] (see Fig. 2 caption for the details). The open symbols represent results from the numerical calculations (squares: magnetization pointing +z, circles: magnetization along -z) whereas the solid/dashed lines correspond to the analytical results. As evident, the analytical solutions agree well with the numerical results.

The x, y, z components of the current induced effective field are shown in Fig. 2(c) and 2(e) when the in-plane field is swept along x- and y-axis, respectively. We use $\hat{p} = (0,1,0)$, $a_J$=3 Oe and $b_J$=3 Oe. The difference of $\Delta H_{X,Y,Z}$ in (c) and (e) is due to the difference in the magnetization azimuthal angle: $\varphi_0 \sim 0$ for (c) and $\varphi_0 \sim \pi/2$ for (e). As shown in Fig. 2(c,e), at low magnetic field, one can consider $\Delta H_X$ and $\Delta H_Y$ as $a_J$ and $b_J$, respectively. We test the validity of Eq. (23) by fitting the external field dependence of the first and second harmonic voltages with parabolic and linear functions, respectively, and calculate quantities corresponding to $B_X$ and $B_Y$ (Eq. (22)). Substituting $B_X$ and $B_Y$ into Eq. (23), we obtain $\Delta H_X \sim -3.03$ Oe and $\Delta H_Y \sim 2.99$ Oe, which match well with $a_J$ and $b_J$ used in the numerical calculations (note that $\Delta H_X = -a_J \cos\theta_0$, see Eq. (41)).

### B. In-plane magnetization systems

#### 1. Small field limit

Numerical results of in-plane magnetized systems for the low field limit are shown in Figs. 3. The material parameters used are relevant for in-plane magnetized systems with a small perpendicular magnetic anisotropy ($H_K \sim -4500$ Oe).

Figure 3 shows results when an out of plane external field, slightly tilted ($\theta_H=5$ deg) from the film normal, is applied. The in-plane component of the tilted field is directed along the magnetic easy axis, which is the x-axis here ($\varphi_H=0$ deg, the in-plane anisotropy field ($H_A$) is $\sim -4$ Oe). The open symbols represent the numerical results. The equilibrium magnetization angles ($\theta_0$, $\varphi_0$) are plotted against the slightly tilted out of plane field in Figs. 3(a) and 3(b), respectively. The magnetization direction reverses (Fig. 3(b)) due to the in-plane component of the tilted field. Figure 3(c) and 3(d) show the first and second harmonic Hall voltages whereas Fig. 3(e) and 3(f) display the first and second harmonic longitudinal voltages. The solid lines represent the analytical solutions, which agree well for the Hall voltages (Fig. 3(c, d)) but show a small deviation at low fields for the longitudinal voltages (Fig. 3(e, f)). The deviation is due to the non-linear (higher order) terms in $R_{XX}$ (Eq. (32)).

To reduce contributions from the non-linear terms in $R_{XX}$, we have used $a_J=0.3$ Oe and $b_J=0.3$ Oe to generate the effective field in Fig. 3 ($\hat{p} = (0,1,0)$ is assumed as before). For $a_J=3$ Oe and $b_J=3$ Oe, as used in the calculations shown in Fig. 2, the non-linear terms dominate the harmonic longitudinal voltages: the analytical solutions do not match the numerical calculations in this external field range (the solution shows better agreement when the external field range is expanded). Note that such effect is negligible for the harmonic Hall voltages (Figs. 3(c) and 3(d)) with $a_J=3$ Oe and $b_J=3$ Oe. The resulting components of the current induced effective field ($\Delta H_{X,Y,Z}$) are shown in the inset of Fig. 3(b). One can identify that $\Delta H_Z$ is the ST term, which changes its sign upon magnetization reversal, and $\Delta H_Y$ is the FL term. Since the magnetization lies along the x-axis, $\Delta H_X$ is nearly zero.

We fit the numerically calculated harmonic Hall voltages vs. field with a linear function and use Eq. (30) to estimate $\Delta H_Y$ (see Fig. 3(d) inset for $V_{2\omega}^{-1}$ vs. the external field and the corresponding linear fit). We obtain $\Delta H_Y \sim 0.32$ Oe for both magnetization direction (pointing along +x and −x). This agrees well with $b_J$ used in the numerical calculations.

For the harmonic longitudinal voltages, we use Eq. (38) to obtain $\Delta H_Z$. Fitting the external field dependence of the first and second harmonic signals with parabolic and linear functions, respectively, we find $\Delta H_Z \sim 0.30$ Oe and $\sim -0.30$ Oe for magnetization pointing along +x and −x, respectively. Although these values match that of $a_J$, it should be noted that non-linear effects start to take place when $a_J$ and $b_J$ becomes large. Since the non-linear terms become apparent when the external field is small, it is desirable to fit the curvature and slope of $V_\omega$ and $V_{2\omega}$, respectively, at higher fields (but much smaller than $H_K$).

### 2. Large range field sweep

We next show the validity of solutions (40a) and (40b) by comparing them to numerical calculations. Solid symbols of Fig. 4 show results from the numerical calculations. The external field is directed nearly along the film normal ($\theta_H$=5 deg) and we apply large enough field to force the magnetic moments to point out of plane. The symbols in Fig. 4(a) and 4(b) show the external field dependence of the equilibrium polar angle of the magnetization and the first harmonic voltage $V_\omega$. The solid line in Fig. 4(a) is calculated from the numerical data shown in Fig. 4(b) using Eq. (40a). Here we take the value of $\Delta R_A \Delta I$ used in the numerical calculations to obtain $\theta_0$ (in experiments, we need to estimate $\Delta R_A \Delta I$ by measuring $V_\omega$ vs. $H$ at $\theta_H=0$ deg). Figures 4(c) and 4(d) show the second harmonic voltage ($V_{2\omega}$) as a function of the external field for $a_J=3$ Oe, $b_J=3$ Oe and $a_J=3$ Oe, $b_J=-3$ Oe, respectively. The resulting x, y, z components of the effective field are shown in Figs. 4(e) and 4(f): $\hat{p} = (0,1,0)$ as before.

As described in Fig. 3(d), the $\sim 1/H$ dependence of $V_{2\omega}$ at low field is due to contribution from the FL term ($\Delta H_Y$, or $b_J$) via the planar Hall effect. $V_{2\omega}$ reverses its sign at low field when Fig. 4(c) and 4(d) are compared, which is due to the difference in the sign of $b_J$. In addition, one can observe a shoulder and a hump-like feature in Figs. 4(c) and 4(d), respectively, when $H\sim\pm|H_K|$ ($H_K \sim -4600$ Oe in Fig. 4). Such feature is due to the first term in Eq. (40b), which is related to the presence of the ST term ($a_J$), that dominates over the second term when $H\sim\pm|H_K|$. We substitute $H_K$, $\theta_H$, $\varphi_H$, $\Delta R_A \Delta I$, $\xi$ used in the numerical calculations and $\theta_0$ obtained from $V_\omega$ (Fig. 4(b)) to Eq. (40b) and fit $V_{2\omega}$ with $a_J$ and $b_J$ as fitting parameters. The results are shown by the solid lines in Fig. 4(c) and 4(d), which agree well with the numerical calculations. The fitting results give $a_J \sim 2.90$ Oe and $b_J \sim 3.08$ Oe for Fig. 4(c) and $a_J \sim 3.10$ Oe and $b_J \sim -3.08$ Oe for Fig. 4(d) that are not far from the nominal values used in the numerical calculations. Thus these results show that one can estimate the spin transfer ($a_J$) and the field like ($b_J$) terms for in-plane magnetized samples by sweeping the field along the film normal direction.

**V. Experimental results**

We study two different film structures to evaluate the current induced effective field using the formula provided above. In particular, we show results from in-plane magnetized samples as this system has not been evaluated in detail previously. The two film structures are: Sub|3 Pt|1

CoFeB|2 MgO|1 Ta and Sub|10 CuIr|$t_{CoFeB}$ CoFeB|2 MgO|1 Ta. The thickness ($t_{CoFeB}$) of the CoFeB layer in the latter film varies from ~1-3 nm: here we use $t_{CoFeB}$~1.3 nm. Both samples are annealed at 300 °C for one hour. The CuIr underlayer film[30] was annealed under the application of in-plane field (4 kOe) along the x-axis in Fig. 1. No field was applied during the annealing for the Pt underlayer film. Both underlayers (Pt and CuIr) have been reported to possess the same sign of the spin Hall angle ($\theta_{SH}$): $\theta_{SH}$~0.01-0.07 for Pt[31-34] and $\theta_{SH}$~0.02-0.03 for CuIr.[30,35] For the CuIr underlayer films the current induced effective field has been estimated using spin torque switching phase diagram measurements,[30] thus giving a reference for comparison. As Pt underlayer typically promotes in-plane anisotropy for CoFeB|MgO, we choose this material as a comparison to CuIr.

Hall bars are patterned from the films using conventional optical lithography and Ar ion etching. 10 Ta|100 Au or 10 Cr|100 Au (units in nm) electrodes are formed by lift-off processes. The width of the Hall bar is ~10 μm: the device structure is similar to that reported previously.[22] We apply a constant amplitude sinusoidal voltage (amplitude: $V_{IN}$, frequency~507.32 Hz) to the Hall bar. Since the Hall bar resistance ($R_{XX}$) shows little dependence on the applied voltage, the source can be considered as a constant amplitude sinusoidal current ($\Delta I$) source. Lock-in amplifiers are used to measure the in-phase first and the out of phase second harmonic voltages simultaneously.

### A. Pt|CoFeB|MgO

Figure 5(a) shows in-plane and out of plane magnetization hysteresis loops measured using vibrating sample magnetometry (VSM). The easy axis is oriented along the film plane and $H_K$~−2000 Oe. Owing to the perpendicular magnetic anisotropy originating at the CoFeB|MgO interface[36], $H_K$ is significantly reduced in magnitude from $4\pi M_S$. Figure 1(b) shows the first

harmonic voltage plotted as a function of the external field swept along different directions, i.e. $\theta_H \sim 0$ deg (black squares), $\theta_H \sim -9$ deg (red circles) and $\theta_H \sim +9$ deg (blue triangles). The in-plane component of the external field is directed along the x-axis (i.e. $\varphi_H \sim 0$). The estimated $H_K$ for $\theta_H \sim 0$ deg more or less agrees with that of the VSM measurements (Fig. 5(a)).

The field dependence of the second harmonic voltages are shown in Fig. 5(c-e) for the three field tilt angles ($\theta_H \sim 0, 9, -9$ deg). When the tilt angle is non-zero, we obtain a curve similar to that of the analytical and numerical calculations shown in Fig. 4(d). Note that the sign of $V_{2\omega}$ reverses when the tilt direction ($\theta_H$) is reversed (see Eq. (40b)). The 0 degree tilt curve should ideally be constant with the external field, however, multi-domain formation at low fields and/or a small misalignment of the field may create features as shown in Fig. 5(d). We use Eq. (40b) to fit $V_{2\omega}$ vs. field for non-zero $\theta_H$: exemplary fitting results are shown by the solid lines in Fig. 5(c) and 5(e). The fitting parameters are $a_J$, $b_J$ and a slight offset voltage, which is typically less than ~1 µV. The obtained $a_J$ and $b_J$ are shown in Fig. 5(f) and 5(g) as a function of the input voltage amplitude $V_{IN}$ for $\theta_H \sim 9$ deg. To show the magnetization direction dependence of the ST and FL terms, $a_J$ and $b_J$ are converted to $\Delta H_Z$ and $\Delta H_Y$, respectively, when the magnetization points along the x-axis. For the ST term, we calculate $\Delta H_Z$ using Eq. (41): $\Delta H_Z = a_J \cos\varphi_H$ when the moments lay within the film plane, i.e. $\theta_0 \sim \pi/2$. Eq. (41) also states that $\Delta H_Y = b_J$.

Both components of the effective field, $\Delta H_Z$ and $\Delta H_Y$, linearly scales with $V_{IN}$. The slope of $\Delta H_{Z(Y)}$ vs. $V_{IN}$ is plotted in Figs. 5(h) and 5(i) as a function of the field tilt angle $\theta_H$ for the ST and FL terms, respectively. The corresponding effective field if a current density of $1 \times 10^8$ A/cm$^2$ were to flow in the Pt underlayer is shown in the right axis. $\Delta H_Z \sim -57$ Oe when both the magnetization and the current are directed along +x; the direction of $\Delta H_Z$ is consistent with the sign of the spin Hall angle associated with the Pt underlayer.[31-34] For $\Delta H_Y$, we show the size of the Oersted field generated when the same magnitude of current is passed through the Pt underlayer. Figure 5(i) shows that $\Delta H_Y$ due to the spin-orbit torque is directed along +y when current is passed along the +x axis and is opposite to that of the Oersted field. Its magnitude is the difference between the measured $\Delta H_Y$ (~29 Oe) and the estimated Oersted field (~−18 Oe), which is ~47 Oe. This is comparable to that of the ST term ($\Delta H_Z$), however the FL term is directed in way that the torque associated with it opposes that of the ST term for current induced magnetization switching,[27] i.e. the FL term points opposite to the incoming electron's spin direction (assuming that the spin Hall effect is the source of the current induced effective field). This seems to be a common feature in magnetic bilayer system,[22-25, 30] in contrast to the ST and FL torques found in MTJs in which the ST and FL torques typically work in sync to switch the direction of magnetization.[37-39] We assume that the angular dependence of the effective field reported recently[23] will be small in the tilt angle range explored here.

### B. CuIr|CoFeB|MgO

Similar experiments are carried out for the CuIr underlayer films. The field dependence of the first harmonic voltage is shown in Fig. 6(a). $\Delta R_A \Delta I$ is smaller than that of the Pt underlayer films since the shunt current through the underlayer is larger. The second harmonic voltages are plotted against the field in Figs. 6(b) and 6(c) for two field tilt angles, $\theta_H$~−9 deg and 9 deg, respectively. In contrast to what have been observed for the Pt underlayer films, here we observe a shoulder instead of a hump at $H$~±$|H_K|$, a feature similar to that obtained by the analytical and numerical calculations shown in Fig. 4(c). Again we use Eq. (40b) to fit the $V_{2\omega}$ vs. $H$ curve to estimate $a_J$ and $b_J$. The fitting results are shown in Figs 6(d) and 6(e) for the ST and FL terms, respectively. Again, we convert $a_J$ and $b_J$ to $\Delta H_Z$ and $\Delta H_Y$ to show the magnetization direction dependence of the effective field explicitly. For magnetization pointing along +x, we find $\Delta H_Z$ to be ~−57 Oe and $\Delta H_Y$ to be ~−21 Oe for current flowing along +x. Taking into account the estimated Oersted field (~−63 Oe) shown by the green dash-dotted line in Fig. 6(e), $\Delta H_Y$ due to the spin orbit torque is ~42 Oe. Both components ($\Delta H_Z$ and $\Delta H_Y$) point along the same direction as that of the Pt underlayer film, which is consistent with the sign of the spin Hall angle of the CuIr underlayer.[30, 3530-35] As with the Pt|CoFeB|MgO system, the FL term points opposite to the incoming electron's spin direction.

We can use the same data set shown in Fig. 6 to test Eq. (30) in evaluating the FL term ($\Delta H_Y$). The inverse of $V_{2\omega}$ is plotted against $H$ in Fig. 7(a) and 7(b) for field tilt angles of ~3 deg and ~9 deg, respectively. A small offset voltage of ~1 µV is subtracted before taking the inverse of $V_{2\omega}$. To evaluate $\partial(1/V_{2\omega})/\partial H$ in Eq. (30), we fit $1/V_{2\omega}$ vs. $H$ in the low field regime with a linear function: the fitting results are shown by the solid lines in Fig. 7(a) and (7b). It is evident that $1/V_{2\omega}$ deviates from a linear function of $H$ at smaller fields for larger $\theta_H$. It is thus preferable to limit $\theta_H$ to a smaller value when performing this analysis.

Equation (30) is used to evaluate ($\Delta H_Y$). As described above, the field dependence of the first harmonic voltage can be used to estimate $H_K$. The low field part ($|H| \lesssim $~1000 Oe) of $V_\omega$ is fitted with a linear function to obtain $\partial V_\omega / \partial H$. Substituting $\partial V_\omega / \partial H$, $\partial(1/V_{2\omega})/\partial H$, $\theta_H$ and the same $H_K$ and $\xi$ values used in Fig. 6 into Eq. (30), we evaluate $\Delta H_Y$ at different $\theta_H$. The input voltage amplitude ($V_{IN}$) dependence of $\Delta H_Y$ is plotted in Fig. 7(c) for $\theta_H$~3 deg. The squares and circles represent the respective $\Delta H_Y$ value when the external field ($H$) is positive and negative (the corresponding x-component of the magnetization ($M_X$) is positive and negative, respectively). The slope of $\Delta H_Y$ vs. $V_{IN}$ is shown against $\theta_H$ in Fig. 7(d). Although the inverse of $V_{2\omega}$ shows a non-linear $H$ dependence at higher fields for larger $\theta_H$ (e.g. Fig. 7(b)), $\Delta H_Y/V_{IN}$ shows a rather small $\theta_H$ dependence. The mean value of $\Delta H_Y/V_{IN}$ is ~−41 Oe (for current density of $1\times10^8$ A/cm$^2$ flowing through the underlayer), which is larger than that shown in Fig. 6(e). Taking into account the Oersted field (~−63 Oe), $\Delta H_Y$ due to the spin orbit torque amounts to ~22 Oe.

We compare the size of the ST and FL terms with those reported previously[30] using spin torque switching phase diagram measurements: the ST and the FL effective fields are reported to be ~−51 Oe and ~14 Oe, respectively, when $1\times10^8$ A/cm$^2$ of current density flows through the CuIr underlayer. The ST term estimated here (~−57 Oe, Fig. 6(d)) agrees well that of Ref. [30]. However, the difference in the FL term is rather large, particularly for the former analysis using the large field range (~42 Oe, Fig. 6(e)). The latter analysis using the low field regime (~22 Oe, Fig. 7(d)) shows better agreement. It is possible that the recently reported angular dependence (the relative angle between the current and the magnetization direction) of the FL term,[23] which is reported to be much larger than that of the ST term, may influence the fitting of $V_{2\omega}$ for a large field range. The discrepancy of the FL term ($\Delta H_T$) for the two analyses we present here (Eq. (30) and Eq. (40b)) requires further investigation for clarification.

Finally, we note that it is difficult to evaluate $\Delta H_Y$ using the small field analysis (Eq. (30)) for the Pt underlayer films as the inverse of $V_{2\omega}$ diverges at $|H|\sim\pm|H_K|/2$ and influences the linear fitting of the low field regime. Similarly, since CoFeB shows small anisotropic magnetoresistance, the second harmonic longitudinal voltage is too small to resolve its field dependence. We thus conclude that fitting the second harmonic voltage for a large field range with Eq. (40a) and (40b) provides a better solution in evaluating the current induced effective field for in-plane magnetized materials than using the low field second harmonic Hall and longitudinal voltages (Eqs. (30) and (38)).

## VI. Conclusion

We have derived analytical formulas that describe the adiabatic (low frequency) harmonic Hall and longitudinal voltages measurements when current induced spin orbit torques develop in magnetic heterostructures. We treat both out of plane and in-plane magnetized samples, taking into account the anomalous and planar Hall effects for the Hall voltage measurements and the anisotropic mangnetoresistance for the longitudinal voltage measurements. The derived forms are compared to numerical calculations using a macrospin model and show good agreement. We experimentally characterize two different in-plane magnetized systems, Pt|CoFeB|MgO and CuIr|CoFeB|MgO, and apply the developed formula to evaluate the effective field in each system. The effective field obtained for the latter system shows relatively good agreement with that evaluated using a spin torque switching phase diagram measurements. Utilizing the harmonic voltage measurements can help gaining solid understanding of the spin orbit torques, which is key to the development of ultrathin magnetic heterostructures for advanced storage class memories and logic devices.

## Acknowledgements

M. H. thanks Kevin Garello and Kyung-Jin Lee for fruitful discussions which stimulated this work. We acknowledge Jaivardhan Sinha for sample preparation and Seiji Mitani for helpful discussions. This work was partly supported by the Grant-in-Aid (25706017) from MEXT and the FIRST program from JSPS.

**References**

1. J. E. Hirsch, Phys. Rev. Lett. **83**, 1834 (1999).
2. S. F. Zhang, Phys. Rev. Lett. **85**, 393 (2000).
3. Y. A. Bychkov and E. I. Rashba, J. Phys. C **17**, 6039 (1984).
4. V. M. Edelstein, Solid State Commun. **73**, 233 (1990).
5. J. C. Slonczewski, J. Magn. Magn. Mater. **159**, L1 (1996).
6. L. Berger, Phys. Rev. B **54**, 9353 (1996).
7. S. Zhang and Z. Li, Phys. Rev. Lett. **93**, 127204 (2004).
8. A. Manchon and S. Zhang, Phys. Rev. B **78**, 212405 (2008).
9. T. Fujita, M. B. A. Jalil, S. G. Tan and S. Murakami, J. Appl. Phys. **110**, 121301 (2011).
10. K. W. Kim, S. M. Seo, J. Ryu, K. J. Lee and H. W. Lee, Phys. Rev. B **85**, 180404 (2012).
11. X. Wang and A. Manchon, Phys. Rev. Lett. **108**, 117201 (2012).
12. D. A. Pesin and A. H. MacDonald, Phys. Rev. B **86**, 014416 (2012).
13. E. van der Bijl and R. A. Duine, Phys. Rev. B **86**, 094406 (2012).
14. P. M. Haney, H. W. Lee, K. J. Lee, A. Manchon and M. D. Stiles, Phys. Rev. B **87**, 174411 (2013).
15. I. M. Miron, K. Garello, G. Gaudin, P. J. Zermatten, M. V. Costache, S. Auffret, S. Bandiera, B. Rodmacq, A. Schuhl and P. Gambardella, Nature **476**, 189 (2011).
16. L. Liu, C.-F. Pai, Y. Li, H. W. Tseng, D. C. Ralph and R. A. Buhrman, Science **336**, 555 (2012).
17. I. M. Miron, T. Moore, H. Szambolics, L. D. Buda-Prejbeanu, S. Auffret, B. Rodmacq, S. Pizzini, J. Vogel, M. Bonfim, A. Schuhl and G. Gaudin, Nat. Mater. **10**, 419 (2011).
18. K.-S. Ryu, L. Thomas, S.-H. Yang and S. Parkin, Nat. Nanotechnol. **8**, 527 (2013).
19. U. H. Pi, K. W. Kim, J. Y. Bae, S. C. Lee, Y. J. Cho, K. S. Kim and S. Seo, Appl. Phys. Lett. **97**, 162507 (2010).
20. I. M. Miron, G. Gaudin, S. Auffret, B. Rodmacq, A. Schuhl, S. Pizzini, J. Vogel and P. Gambardella, Nat. Mater. **9**, 230 (2010).
21. T. Suzuki, S. Fukami, N. Ishiwata, M. Yamanouchi, S. Ikeda, N. Kasai and H. Ohno, Appl. Phys. Lett. **98**, 142505 (2011).
22. J. Kim, J. Sinha, M. Hayashi, M. Yamanouchi, S. Fukami, T. Suzuki, S. Mitani and H. Ohno, Nat. Mater. **12**, 240 (2013).

23. K. Garello, I. M. Miron, C. O. Avci, F. Freimuth, Y. Mokrousov, S. Blugel, S. Auffret, O. Boulle, G. Gaudin and P. Gambardella, Nat. Nanotechnol. **8**, 587 (2013).
24. S. Emori, U. Bauer, S.-M. Ahn, E. Martinez and G. S. D. Beach, Nat Mater **12**, 611 (2013).
25. X. Fan, J. Wu, Y. Chen, M. J. Jerry, H. Zhang and J. Q. Xiao, Nat Commun **4**, 1799 (2013).
26. P. Balaz, J. Barnas and J. P. Ansermet, J. Appl. Phys. **113**, 193905 (2013).
27. S. Zhang, P. M. Levy and A. Fert, Phys. Rev. Lett. **88**, 236601 (2002).
28. J. Z. Sun, Phys. Rev. B **62**, 570 (2000).
29. M. D. Stiles and J. Miltat, in *Spin Dynamics in Confined Magnetic Structures Iii* (Springer-Verlag Berlin, Berlin, 2006), Vol. 101, pp. 225.
30. M. Yamanouchi, L. Chen, J. Kim, M. Hayashi, H. Sato, S. Fukami, S. Ikeda, F. Matsukura and H. Ohno, Appl. Phys. Lett. **102**, 212408 (2013).
31. L. Q. Liu, O. J. Lee, T. J. Gudmundsen, D. C. Ralph and R. A. Buhrman, Phys. Rev. Lett. **109**, 096602 (2012).
32. O. Mosendz, J. E. Pearson, F. Y. Fradin, G. E. W. Bauer, S. D. Bader and A. Hoffmann, Phys. Rev. Lett. **104**, 046601 (2010).
33. M. Morota, Y. Niimi, K. Ohnishi, D. H. Wei, T. Tanaka, H. Kontani, T. Kimura and Y. Otani, Phys. Rev. B **83**, 174405 (2011).
34. L. Liu, R. A. Buhrman and D. C. Ralph, cond-mat., 1111.3702 (2011).
35. Y. Niimi, M. Morota, D. H. Wei, C. Deranlot, M. Basletic, A. Hamzic, A. Fert and Y. Otani, Phys. Rev. Lett. **106**, 126601 (2011).
36. S. Ikeda, K. Miura, H. Yamamoto, K. Mizunuma, H. D. Gan, M. Endo, S. Kanai, J. Hayakawa, F. Matsukura and H. Ohno, Nat. Mater. **9**, 721 (2010).
37. J. C. Sankey, Y.-t. Cui, J. Z. Sun, J. C. Slonczewski, R. A. Buhrman and D. C. Ralph, Nature Phys. **4**, 67 (2008).
38. H. Kubota, A. Fukushima, K. Yakushiji, T. Nagahama, S. Yuasa, K. Ando, H. Maehara, Y. Nagamine, K. Tsunekawa, D. D. Djayaprawira, N. Watanabe and Y. Suzuki, Nat. Phys. **4**, 37 (2008).
39. S. C. Oh, S. Y. Park, A. Manchon, M. Chshiev, J. H. Han, H. W. Lee, J. E. Lee, K. T. Nam, Y. Jo, Y. C. Kong, B. Dieny and K. J. Lee, Nat. Phys. **5**, 898 (2009).

**Figure Captions**

**Fig. 1**. Schematic illustration of the experimental setup. A Hall bar is patterned from a magnetic heterostructure consisting of a non-magnetic metal layer (gray), a ferromagnetic metal layer (blue) and an insulating oxide layer (red). The large gray square is the substrate with an insulating oxide surface. Definitions of the coordinate systems are illustrated together. $\vec{M}$ denotes the magnetization and $\vec{H}$ represents the external field.

**Fig. 2**. (a) Magnetization angle with respect to the film normal ($\theta_0$), (b) first harmonic Hall voltage and (d,f) second harmonic Hall voltage plotted against in-plane external field ($\theta_H$=90 deg). The field is directed along the x-axis ($\varphi_H$=0 deg) for (a,b,d) and along the y-axis ($\varphi_H$=90 deg) for (f). Open symbols show numerical calculations using the macrospin model. Solid/dashed lines represent the analytical solutions: (a) Eq. (17), (b) Eq. (21a), (d,f) Eq. (21b). (c,e) x, y, z component of the effective field used for the numerical calculations. Left (right) panel indicates the effective field when the magnetization is pointing along +z (-z). Parameters used in the numerical calculations: $H_K$=3162 Oe, $H_A$=−6 Oe, $\alpha$=0.01, $\gamma$=17.6 MHz/Oe, $a_J$=3 Oe, $b_J$=3 Oe, $\hat{p}=(0,1,0)$, $\Delta R_A$=1 Ω, $\Delta R_P$=0.1 Ω, $\Delta I$=1 A.

**Fig. 3**. (a) Polar ($\theta_0$) and (b) azimuthal ($\varphi_0$) angles of the magnetization, (c) first and (d) second harmonic Hall voltages ($V_{XY}$), (e) first and (f) second harmonic longitudinal voltages ($V_{XX}$) as a function of a slightly tilted out of plane field ($\theta_H$=5 deg). The in-plane component of the tilted field is directed along the x-axis ($\varphi_H$=0 deg). Open symbols show numerical calculations using

the macrospin model. Solid lines represent the analytical solutions: (a,b) Eq. (24), (c) Eq. (26), (d) Eq. (28), (e) Eq. (37a), (f) Eq. (37b). Inset of (b) shows the x, y, z component of the effective field used for the numerical calculations. Parameters used in the numerical calculations: $H_K$=−4657 Oe, $H_A$=−4 Oe, $\alpha$=0.01, $\gamma$=17.6 MHz/Oe, $a_J$=0.3 Oe, $b_J$=0.3 Oe, $\hat{p}=(0,1,0)$, $\Delta R_A$=1 Ω, $\Delta R_P$=0.1 Ω, $\Delta R_{MR}$=1 Ω, $R_0$=0 Ω, $\Delta I$=1 A.

**Fig. 4**. (a) Polar ($\theta_0$) angle of the magnetization, (b) first and (c,d) second harmonic Hall voltages as a function of a slightly tilted out of plane field ($\theta_H$=5 deg). The in-plane component of the tilted field is directed along the x-axis ($\varphi_H$=0 deg). Solid symbols show numerical calculations using the macrospin model. Solid lines represent the analytical solutions: (a) Eq. (40a), (c,d) Eq. (40b). (e,f) The x, y, z components of the effective field used for the numerical calculations. Parameters used in the numerical calculations: $H_K$=−4561 Oe, $H_A$=−4 Oe, $\alpha$=0.01, $\gamma$=17.6 MHz/Oe, $\hat{p}=(0,1,0)$, $\Delta R_A$=1 Ω, $\Delta R_P$=0.1 Ω, $\Delta R_{MR}$=1 Ω, $R_0$=0 Ω, $\Delta I$=1 A. (c,e) $a_J$=3 Oe, $b_J$=3 Oe, (d,f) $a_J$=3 Oe, $b_J$=−3 Oe.

**Fig. 5**. Results for sub|3 Pt|1 CoFeB|2 MgO|1 Ta (units in nm). (a) Magnetization hysteresis loops: solid and open symbols represent out of plane ($H_\perp$) and in-plane ($H_{//}$) field sweeps, respectively. (b) First harmonic voltage as a function of external field with different tilt angles ($\theta_H$). $V_{IN}$=0.5 V. (c-e) External field dependence of the second harmonic voltage ($V_{2\omega}$) for three different $\theta_H$. $V_{IN}$=3.5 V. The solid lines represent fitting results using Eq. (40b). (f,g) $\Delta H_Z$ (f) and $\Delta H_Y$ (g) obtained from the fitting of $V_{2\omega}$ plotted against the excitation voltage amplitude

($V_{IN}$) for $\theta_H$~9 deg. (h,i) $\theta_H$ dependence of the effective field per unit excitation voltage amplitude for the spin transfer term, $\Delta H_Z$ (h) and the field-like term, $\Delta H_Y$ (i). The right axis show the corresponding effective field if a current density of $1 \times 10^8$ A/cm$^2$ were applied to the underlayer (3 Pt). Square and circular symbols in (f-i) represent the corresponding effective field for magnetization pointing along +x and –x axes, respectively. The green dashed line in (i) show the calculated Oersted field if all the current flows into the underlayer. The field is calculated for 1 nm above the top surface of the Pt underlayer.

**Fig. 6**. Results for sub|10 CuIr|1 CoFeB|2 MgO|1 Ta (units in nm). (a) First harmonic voltage as a function of external field with different tilt angles ($\theta_H$). $V_{IN}$=0.5 V. (b,c) External field dependence of the second harmonic voltage ($V_{2\omega}$) for $\theta_H$~−9 (b) and 9 deg (c). $V_{IN}$=3.5 V. The solid lines represent fitting results using Eq. (40b). (d,e) $\theta_H$ dependence of the effective field per unit excitation voltage amplitude for the spin transfer term, $\Delta H_Z$ (d) and the field-like term, $\Delta H_Y$ (e). The right axis show the corresponding effective field if a current density of $1 \times 10^8$ A/cm$^2$ were applied to the underlayer (10 nm CuIr). Square and circular symbols represent the corresponding effective field for magnetization pointing along +x and –x axes, respectively. The green dashed line in (e) show the calculated Oersted field if all the current flows into the underlayer. The field is calculated for 1 nm above the top surface of the CuIr underlayer.

**Fig. 7**. Results for sub|10 CuIr|1 CoFeB|2 MgO|1 Ta (units in nm). (a,b) External field dependence of the inverse of the second harmonic voltage ($1/V_{2\omega}$) for $\theta_H$~3 (b) and 9 deg (c). The solid lines represent linear fitting. $V_{IN}$=3.5 V. (c) $\Delta H_Y$ obtained using Eq. (30) plotted

against the excitation voltage amplitude ($V_{IN}$) for $\theta_H$~3 deg. (d) $\theta_H$ dependence of the effective field per unit excitation voltage amplitude for the field-like term, $\Delta H_Y$. The right axis show the corresponding effective field if a current density of $1 \times 10^8$ A/cm$^2$ were applied to the underlayer (10 CuIr). Square and circular symbols represent the corresponding effective field for magnetization pointing along +x and −x axes, respectively. The green dashed line in (d) show the calculated Oersted field if all the current flows into the underlayer (same as in Fig. 6(e)).

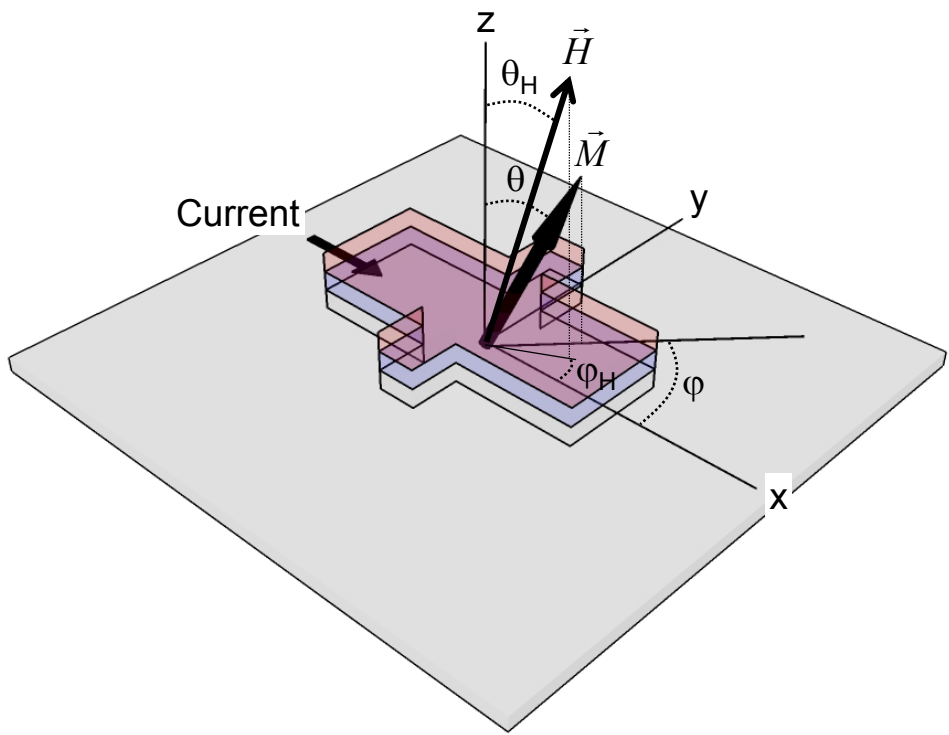

**Fig. 1**

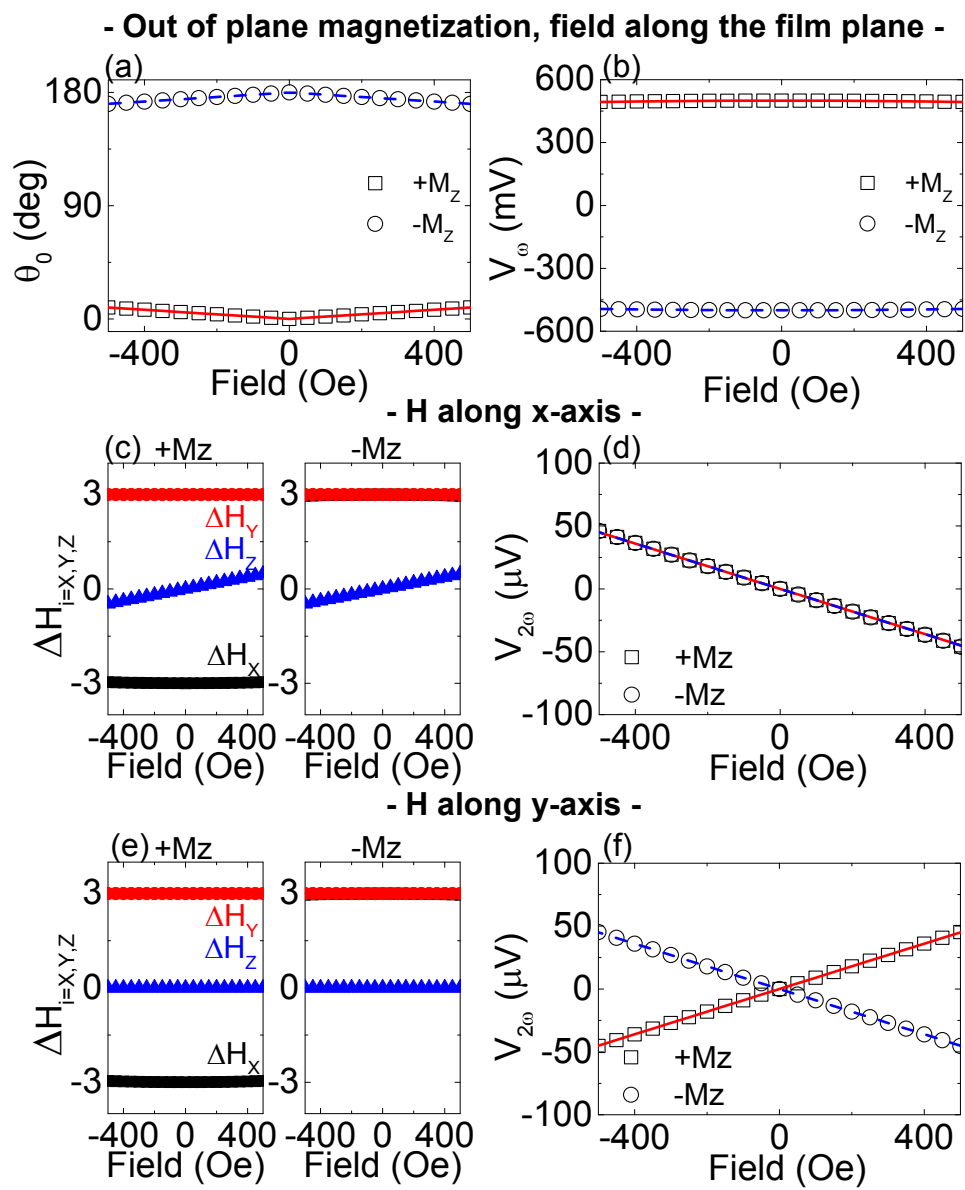

**Fig. 2**

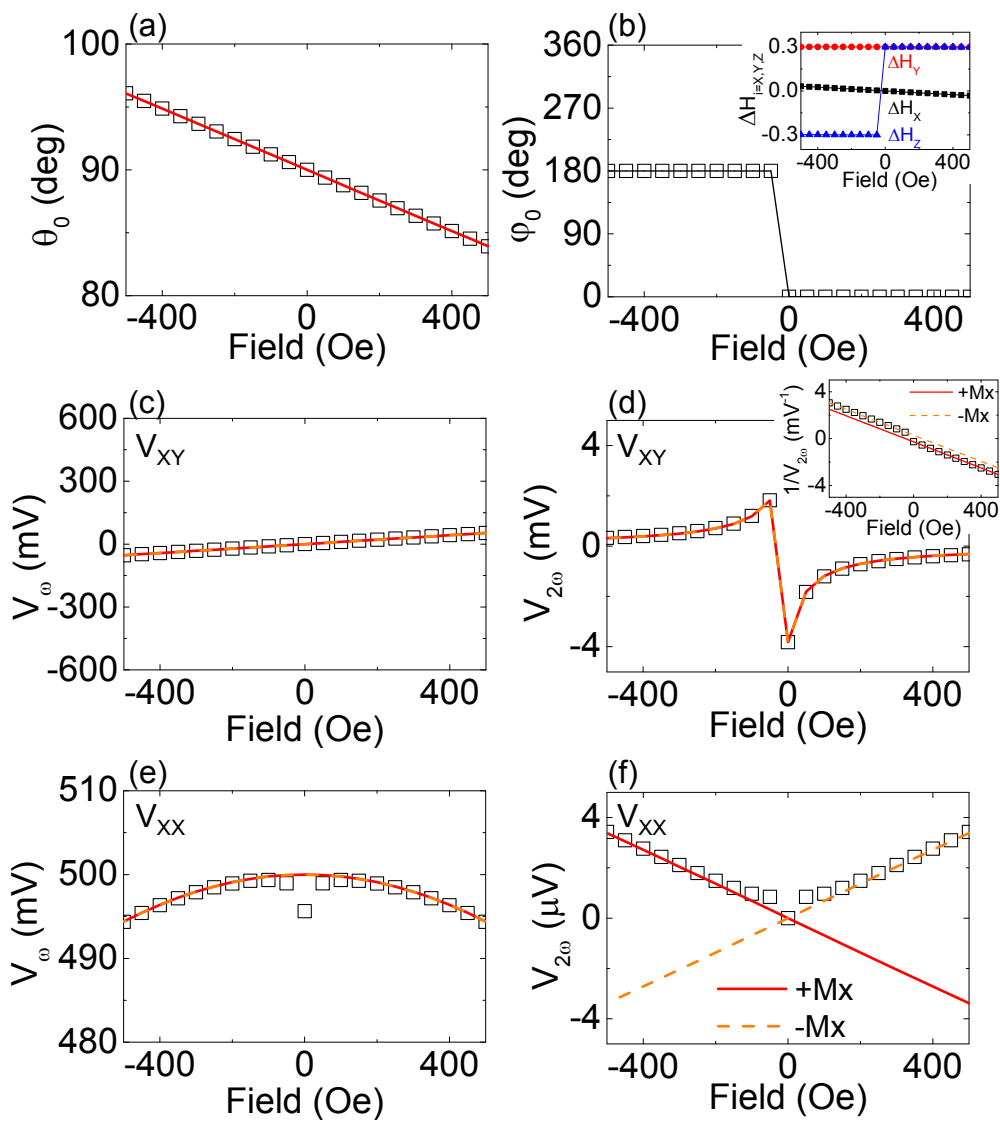

**Fig. 3**

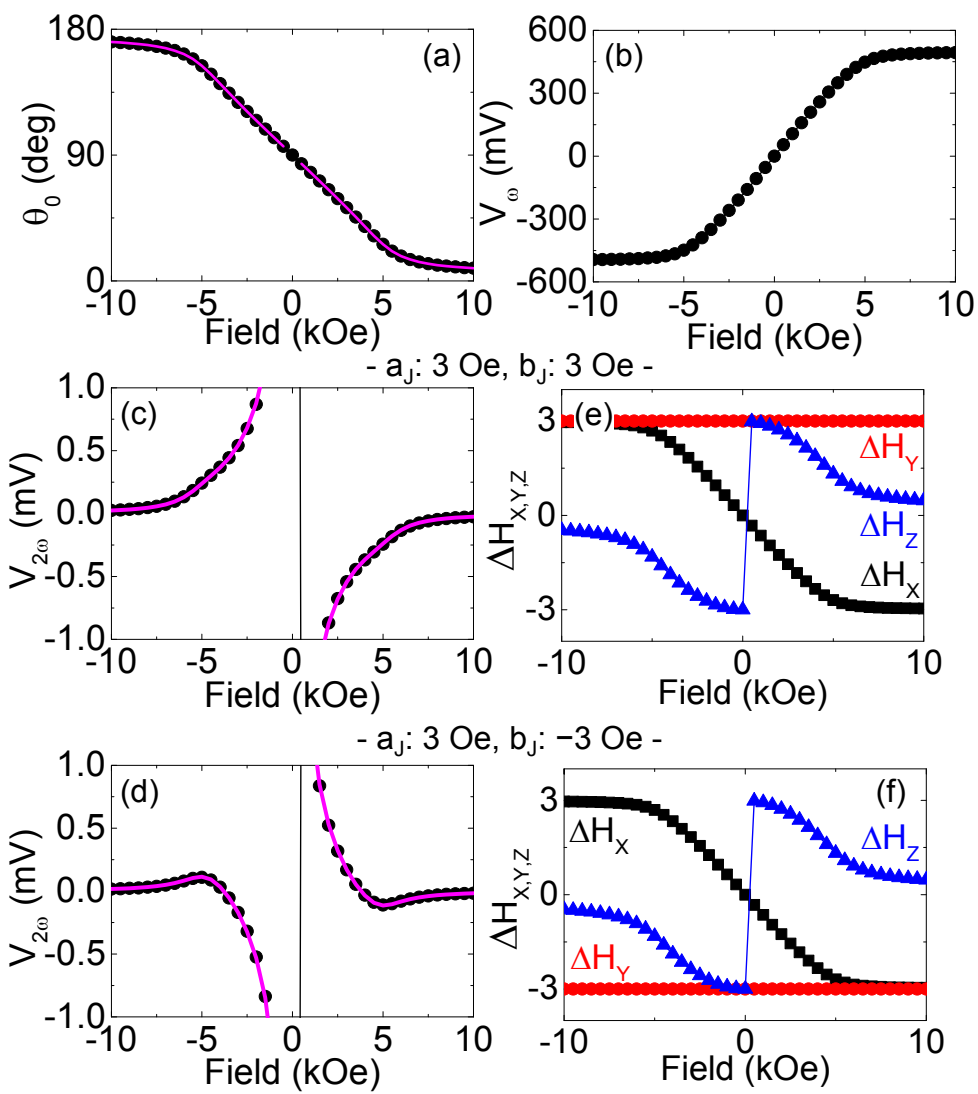

Fig. 4

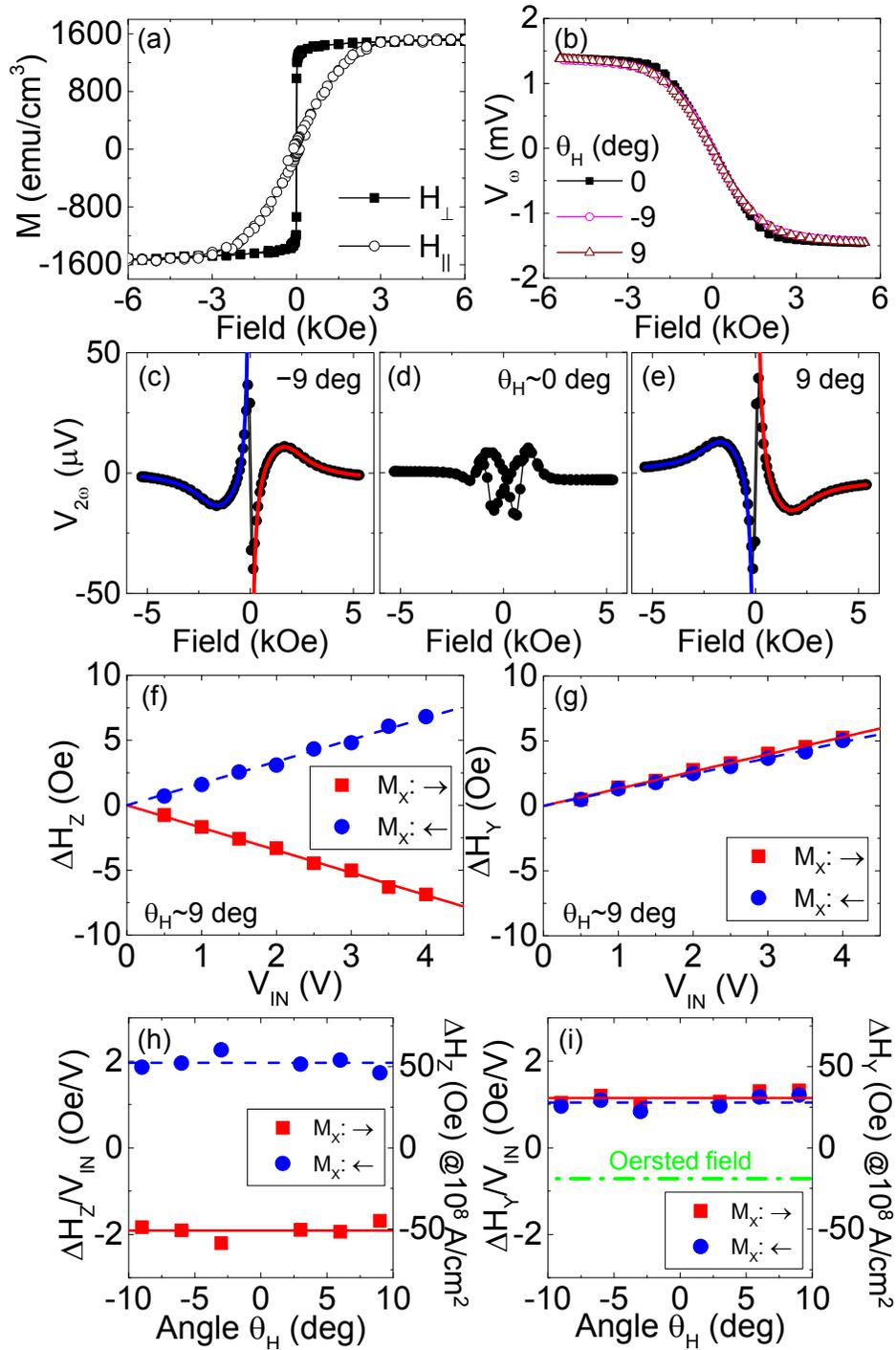

**Fig. 5**

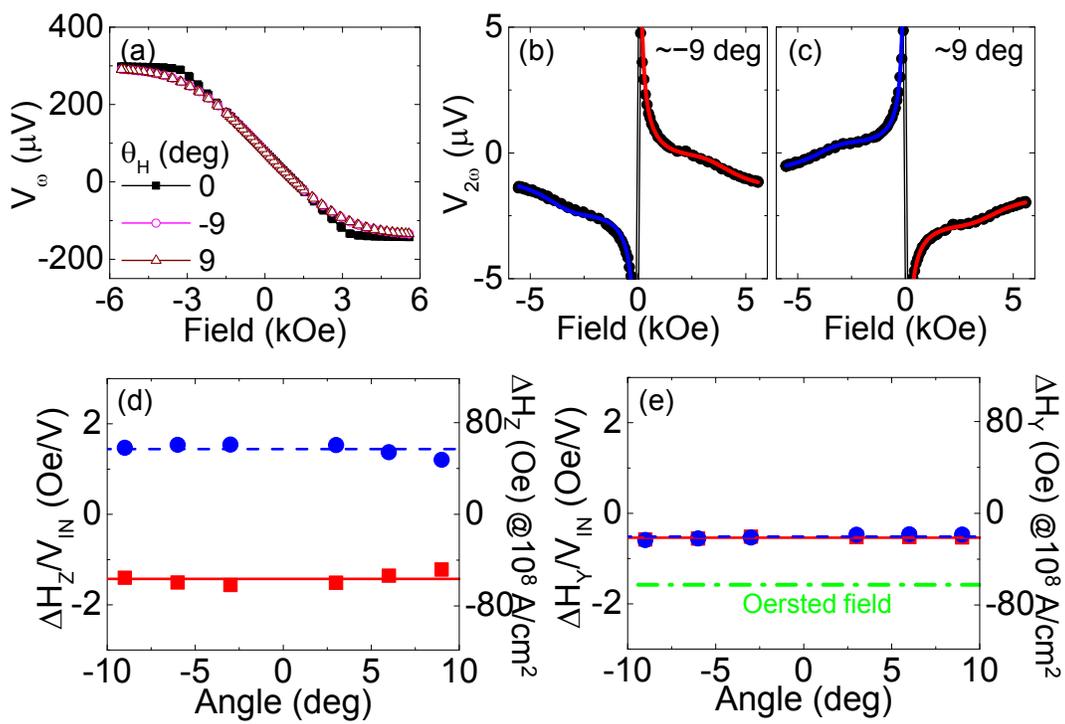

Fig. 6

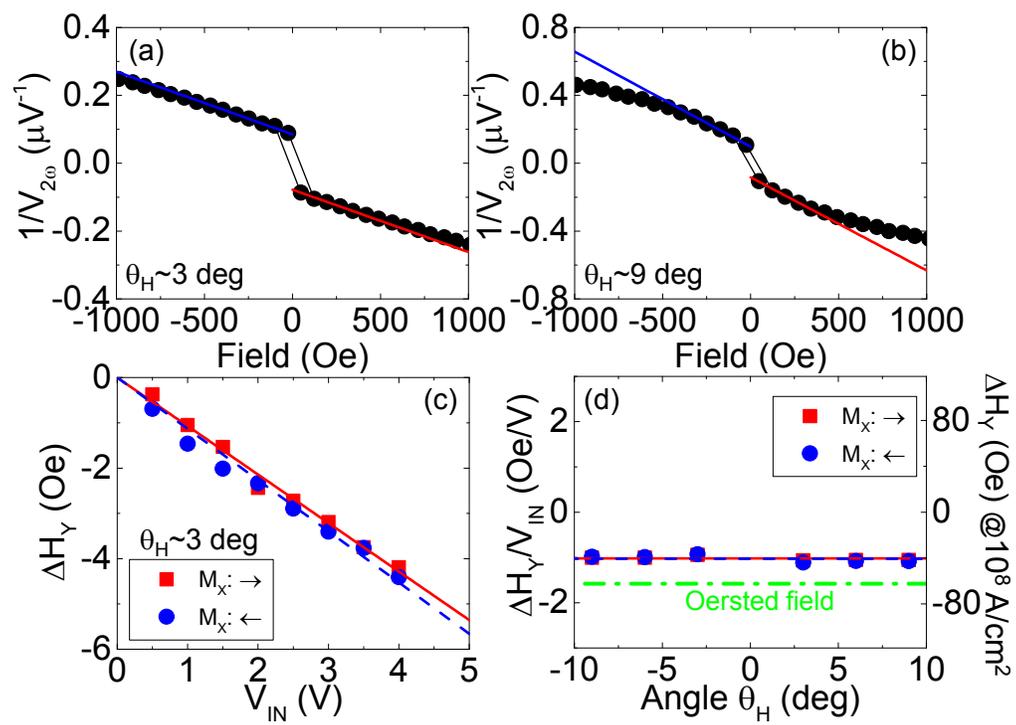

**Fig. 7**